\documentclass[fleqn,usenatbib]{mnras}

\usepackage{newtxtext,newtxmath}

\usepackage[T1]{fontenc}

\DeclareRobustCommand{\VAN}[3]{#2}
\let\VANthebibliography\thebibliography
\def\thebibliography{\DeclareRobustCommand{\VAN}[3]{##3}\VANthebibliography}

\usepackage{graphicx}	
\usepackage{amsmath}	

\usepackage{longtable} 
\usepackage{threeparttable}

\title[Studying LSB structures from annotated deep images]{Low Surface Brightness structures from annotated deep CFHT images: effects of the host galaxy's properties and environment }

\author[E. Sola et al.]{
Elisabeth Sola,$^{1,2}$
Pierre-Alain Duc,$^{2}$
Mathias Urbano,$^{2}$
Felix Richards,$^{3}$
Adeline Paiement,$^{4}$
\newauthor
Michal B\'ilek,$^{5,6,7}$ 
Mustafa K. Y{\i}ld{\i}z,$^{8,9}$
Alessandro Boselli,$^{10}$ 
Patrick C\^{o}t\'{e},$^{11}$ 
Jean-Charles Cuillandre,$^{12}$  
\newauthor
Laura Ferrarese,$^{11}$
Stephen Gwyn,$^{11}$
Olivier Marchal,$^{2}$
Alan W. McConnachie,$^{11}$
Matthieu Baumann,$^{2}$
\newauthor
Thomas Boch,$^{2}$
Florence Durret,$^{13}$
Matteo Fossati,$^{14,15}$
Rebecca Habas,$^{16}$  
Francine Marleau,$^{17}$
\newauthor
Oliver M\"{u}ller,$^{18}$
M\'{e}lina Poulain,$^{19}$
Vasily Belokurov$^{1}$
\\
$^{1}$ Institute of Astronomy, Madingley Rd, Cambridge CB3 0HA, UK \\
$^{2}$ Universit\'e de Strasbourg, CNRS, Observatoire astronomique de Strasbourg (ObAS), UMR 7550, F-67000 Strasbourg, France\\
$^{3}$ Department of Computer Science, Swansea University, UK\\
$^{4}$ Universit\'e de Toulon, Aix Marseille Univ, CNRS, LIS, Marseille, France \\
$^{5}$ Astronomical Observatory of Belgrade, Volgina 7, 11060 Belgrade, Serbia\\
$^{6}$ LERMA, Observatoire de Paris, CNRS, PSL Univ., Sorbonne Univ., 75014 Paris, France\\
$^{7}$ Department of Astrophysics, University of Vienna, T\"urkenschanzstra{\ss}e 17, 1180 Vienna, Austria\\
$^{8}$ Astronomy and Space Sciences Department, Science Faculty, Erciyes University, Kayseri, 38039, Türkiye\\
$^{9}$ Erciyes University, Astronomy and Space Sciences Observatory Applied and Research Center (UZAYB\.{I}MER), 38039, Kayseri, Türkiye\\
$^{10}$ Aix-Marseille Univ., CNRS, CNES, LAM, Marseille, France\\
$^{11}$ Herzberg Astronomy and Astrophysics Research Centre, National Research Council of Canada, Victoria, BC V9E 2E7, Canada \\
$^{12}$ Universit\'e Paris-Saclay, Universit\'e Paris Cit\'e, CEA, CNRS, AIM, 91191, Gif-sur-Yvette, France \\
$^{13}$ Sorbonne Universit\'e, CNRS, UMR 7095, Institut d’Astrophysique de Paris, 98bis Bd Arago, 75014 Paris, France\\
$^{14}$ Dipartimento di Fisica G. Occhialini, Universit\`a degli Studi di Milano-Bicocca, Piazza della Scienza 3, 20126 Milano, Italy \\
$^{15}$ INAF - Osservatorio Astronomico di Brera, via Brera 28, 21021 Milano, Italy \\
$^{16}$ INAF-Astronomical Observatory of Abruzzo, Via Maggini, 64100, Teramo, Italy\\
$^{17}$ Universit\"at Innsbruck, Institut f\"ur Astro- und Teilchenphysik, Technikerstr. 25/8, 6020 Innsbruck, Austria\\
$^{18}$ Institute of Physics, Laboratory of Astrophysics, Ecole Polytechnique F\'{e}d\'{e}rale de Lausanne (EPFL), 1290 Sauverny, Switzerland\\
$^{19}$ Space Physics and Astronomy Research Unit, University of Oulu, P.O. Box 3000, FI-90014, Oulu, Finland\\    
}

\date{Accepted 2025 July 07. Received 2025 June 27; in original form 2025 March 17}

\pubyear{\the\year{}}

\begin{document}
\label{firstpage}
\pagerange{\pageref{firstpage}--\pageref{lastpage}}
\maketitle

\begin{abstract}
Hierarchical galactic evolution models predict that mergers drive galaxy growth, producing low surface brightness (LSB) tidal features that trace galaxies’ late assembly. These faint structures encode information about past mergers and are sensitive to the properties and environment of the host galaxy. We investigated the relationships between LSB features and their hosts in a sample of 475 nearby massive galaxies spanning diverse environments (field, groups, Virgo cluster) using deep optical imaging from the Canada-France-Hawaii Telescope (MATLAS, UNIONS/CFIS, VESTIGE, NGVS). Using \textit{Jafar}, an online annotation tool, we manually annotated tidal features, including 199 tidal tails and 100 streams. Geometric and photometric measurements were extracted to analyse their dependence on galaxy mass, environment, and internal kinematics.
At our surface brightness limit of 29 mag$\,$arcsec$^{-2}$, tidal features contribute 2\% of total galaxy luminosity. They are detected in 36\% of galaxies, with none fainter than 27.8 mag$\,$arcsec$^{-2}$. The most massive galaxies are twice as likely to host tidal debris. Although small-scale interactions increase the frequency of tidal features, the large-scale environment (Virgo cluster versus field/group) does not influence it. An anticorrelation between this frequency and rotational support is found, but may reflect the mass-driven effect. We release our database of annotated features for deep learning applications. Our findings confirm that galaxy mass is the dominant factor influencing tidal feature prevalence, consistent with hierarchical formation models.

\end{abstract}

\begin{keywords}
Galaxies: interactions -- Galaxies: evolution -- Methods: data analysis
\end{keywords}



\section{Introduction}
In the framework of hierarchical models of galaxy evolution, galaxies are expected to be assembled through mergers and smooth accretion of cold intergalactic gas. The dominant growth mode is a function of time, environment, mass and morphological type \citep[e.g.,][]{white_et_al_1991,Kauffman_et_al_1993,Cole_et_al_2000,Baugh_et_al_2003,Keres_et_al_2005,Daddi_et_al_2007,Ocvirk_et_al_2008,Dekel_et_al_2009,Forster_Schreiber_2009,Agertz_et_al_2009,Stringer_and_Benson_2007,Genel_et_al_2010,Lhuillier_et_all_2012}.
Massive early-type galaxies (ETGs) are most likely formed in a two-phase scenario first dominated by smooth accretion of gas and gas-rich mergers and then by minor mergers \citep[e.g.][]{Oser_et_al_2010,Thomas_et_al_1999,Naab_et_al_2009} or more rarely from the merger of two spirals \citep[e.g.,][]{Toomre_1977,Barnes_1988}. In contrast, the formation of late-type galaxies (LTGs) is likely smoother, mainly through accretion of cold intergalactic gas \citep[e.g.,][]{Sancisi_et_al_2008,Bilek_et_al_2022b}.
The environment in which galaxies reside also drives their evolution, which has been confirmed observationally and with numerical simulations \citep[e.g.,][]{Gunn_and_Gott_1972, Oemler_1974, Dressler_1980, Larson_et_al_1980, Byrd_and_Valtonen_1990, Moore_et_al_1996, Bekki_1999, Goto_et_al_2003, Mihos_2003, Hester_2006, Kawata_et_al_2007, Berrier_et_al_2008, Boselli_et_al_2022}.

Mergers and interactions between galaxies  lead to various modifications of galaxies' properties, such as their  morphology, kinematics, chemical composition, stellar metallicity and age gradients \citep[e.g.,][]{Balcells_et_al_1990,Di_Matteo_et_al_2009,Bois_et_al_2011,Schulze_et_al_2017,Amorisco_et_al_2017,Bilek_et_al_2022b}.
Such interactions produce tidal structures that extend around galaxies. The nature, morphology, and lifetime of these features highly depend on the age, mass, type, and impact parameters of the collisions and mergers. The mass ratio between the interacting galaxies, the inclinations of their orbits and differences in velocities are essential parameters \citep[e.g.,][]{Pop_et_al_2018, Mancillas_et_al_2019, shells_distance2}. For instance, tidal tails are likely formed during major mergers \citep[e.g.,][]{Arp_1966, Toomre_and_Toomre_1972}, but can also arise from non-merging fly-bys \citep[e.g.,][]{Toomre_and_Toomre_1972,Sinha_et_al_2012,Duc_and_Renaud_2011}. Streams are presumably formed during minor mergers  \citep[e.g.,][]{Bullock_and_Johnston_2005, Belokurov_et_al_2006}, and shells are typically formed during intermediate-mass radial mergers \citep[e.g.,][]{Quinn_1984,Prieur_1990}. Hence, the different types of tidal structures hold crucial clues about the assembly history of galaxies.
Numerical simulations in cosmological contexts help to interpret the prevalence and frequency of tidal features and extended stellar haloes 
\citep[e.g.,][]{Hernquist_et_al_1987,Johnston_et_al_1999,Ibata_et_al_2001a,Johnston_et_al_2008,Cooper_et_al_2010,Lux_et_al_2013,Ebrova_2013,Hendel_and_Johnston_2015,Thomas_et_al_2017,Pop_et_al_2018,Hood_et_al_2018,Mancillas_et_al_2019,Bilek_et_al_2022b}.

Tidal features are also affected by the environment in which they reside. In the field, without external perturbations, stripped tidal debris remains loosely bound and can be visible for several Gyr \citep[e.g.,][]{Mancillas_et_al_2019}. In clusters, they are the result of competing processes that can either generate or erase them \citep[e.g.,][]{Byrd_and_Valtonen_1990, Moore_et_al_1996, Mihos_2003,Gnedin_2003}. The diffuse debris contributes to the intracluster light, which is made of stars no longer bound to any galaxy \citep[e.g.,][]{Rudick_et_al_2006,Conroy_et_al_2007,Rudick_et_al_2009,Montes_2019}.

While a wealth of tidal features have been observed in the Local Group \citep[e.g.,][]{Ibata_et_al_2001,Belokurov_et_al_2006,McConnachie_et_al_2009,Martin_et_al_2014}, their detection is challenging for objects located further away that are no longer resolved into stars. The diffuse, Low Surface Brightness (LSB), light they emit is difficult to detect. Unveiling the faint diffuse tidal features requires LSB optimized deep imaging as well as dedicated reduction pipelines. Recent progress in CCD cameras, observing strategies and data processing techniques enabled us to reach the LSB regime ($>25$ mag$\,$arcsec$^{-2}$) with appropriate background subtraction \citep[e.g.,][]{elixir,Borlaff_2019}. Deep observations of several hundreds of square degrees have been conducted using a large number of telescopes, from small-aperture telescopes \citep[e.g.,][]{Martinez_Delgado_et_al_2010,Abraham_and_van_Dokkum_2014,Spitler_et_al_2019,Mosenkov_et_al_2020} to large ones \citep[e.g.,][]{Mihos_et_al_2005, Ferrarese_et_al_2012,Boselli_et_al_2018,Duc_2020,Venhola_et_al_2017,Iodice_et_al_2021,Alabi_et_al_2021,Jackson_et_al_2021}. The upcoming Euclid and Rubin/LSST   surveys are about to extend the realm of the LSB Universe even further, with thousands of square degrees covered \citep[e.g.,][]{Borlaff_2022,Martin_et_al_2022}.

Different methods have been developed to identify tidal features in deep images. Among them, the basic visual inspection of images conducted by one or more collaborators  remains popular \citep[e.g.,][]{Tal_et_al_2009,Atkinson_Abraham_Ferguson_2013,Morales_et_al_2018,Hood_et_al_2018,Muller_et_al_2019,Kluge_et_al_2020,Bilek_et_al_2020,Jackson_et_al_2021}. For samples of thousands of galaxies, simple classification tasks can also be done by a crowd of citizen scientists \citep[e.g.,][]{Casteels_et_al_2012}. However, as the sample size increases, visual detection is no longer possible, and it is necessary to rely on automated techniques \citep[e.g.,][]{Pawlik_et_al_2016,Mantha_et_al_2019,Kado-Fong_et_al_2018}. Non-parametric methods like the Gini-M$_{20}$ parameter \citep[e.g.,][]{Abraham_et_al_2003,Lotz_2004} or the CAS system \citep[e.g.,][]{Conselice_et_al_2003,Conselice_2009} are also used to characterise the host galaxies. However, they are weakly sensitive to the presence of the LSB structures of interest. Deep learning and convolutional neural networks (CNNs) provide promising approaches  \citep[e.g.,][]{Pearson_et_al_2019,Walmsley_et_al_2018,Martin_et_al_2020,Bickley_et_al_2021} but remain to be thoroughly tested on complex images.

Several limitations arise from the aforementioned studies. First, automated methods still need to provide subtle classifications and distinctions between the different types of tidal features. Furthermore, quantitative measurements of the properties of LSB tidal features that are obtained consistently over large samples of galaxies are still missing. However, they are needed to constrain specific models of galactic evolution. So far, most studies focused on the census of tidal debris have performed qualitative analyses \citep[e.g.,][]{Tal_et_al_2009,Miskolczi_et_al_2011,Casteels_et_al_2012,Morales_et_al_2018,Bilek_et_al_2020}. Those that have provided quantitative results relied  on limited samples \citep[e.g.,][]{Martinez-Delgado_et_al_2021,Huang_and_Fan_2022}.

\cite{Sola_et_al_2022} introduced a dedicated annotation tool, \textit{Jafar}, and an associated database, which partially address these difficulties. They enable the retrieval of  quantitative measurements of LSB tidal features through the precise delineation of features superimposed on deep images. This online tool\footnote{\textit{Jafar} relies on developments made by a computer vision team \citep{Richards_et_al_2021,Richards_et_al_2024} and the CDS Aladin team.}  allows contributors to annotate images of  hundreds of galaxies in a user-friendly way.

This paper is a follow-up of a series of studies based on the census of LSB structures  \citep[][]{Duc_et_al_2015,Bilek_et_al_2020,Bilek_et_al_2022b,Sola_et_al_2022} identified in the ATLAS$^{3D}$ \citep{Cappellari_et_al_2011} volume-limited sample of massive nearby galaxies. In particular, \cite{Bilek_et_al_2020, Bilek_et_al_2022b} performed a comprehensive study of tidal features around ETGs. They investigated the origin of the differences in rotational support among ETGs in the field and groups. They found that the kinematics of the ETGs were established early in the formation history of the galaxies. Complementary to that study, and others carried out at higher redshifts \citep[e.g.,][]{Bridge_et_al_2010,Wen_and_Zheng_2016}, we investigate the late galactic evolution in this paper. We probe the merger history within the past few Gyr for which the LSB features visible today may bring observational constraints. We extend the previous sample of \cite{Bilek_et_al_2020} by including LTGs and the Virgo cluster as an additional environment. We target galaxies undergoing interactions and post-merger systems without knowing whether they host tidal features.

The article is organised as follows. In Section \ref{section:data}, we present the data and galaxy sample, and our methods are explained in Section \ref{section:methods}. We present the general properties of LSB features in Section \ref{section:results-properties_features}. Results concerning the effect of galaxy mass and environment on tidal features are presented in Sections \ref{section:results-effect_galaxy_mass} and \ref{section:results-effect_environment}, respectively, while their joint impact is shown in Section \ref{section:results-joint_impact_mass_env} and the impact of internal kinematics is presented in Section \ref{section:results-effect_internal_kinematics}. We discuss the results and compare them to the literature in Section \ref{section:discussion}, and the conclusions are presented in Section \ref{section:conclusion}.
\section{Sample and data} \label{section:data}

This section describes the data we used to probe galactic growth through LSB tidal features. Briefly, we analyse galaxies extracted from the reference sample compiled by the ATLAS$^{3D}$ collaboration \citep{Cappellari_et_al_2011}, that contains 260 ETGs (the primary target of the ATLAS$^{3D}$ project) and 611 LTGs (from their parent sample). 
Our final galaxy sample is composed of the galaxies from the ATLAS$^{3D}$ reference sample that lie in the footprint of our images. We detail the characteristic of the galaxies catalogues and images below.

ETGs and LTGs from the ATLAS$^{3D}$ reference sample are characterised by being bright (absolute magnitude in $K$-band $M_K<-21.5$ mag), massive (stellar mass $M_\star > 6\times 10^{9} M_\odot$), nearby (distance $D<46$ Mpc), and visible from the Northern hemisphere ($\lvert \delta - 29^{\circ} \rvert< 35^{\circ}  $ and $\lvert b \rvert > 15 ^{\circ}$ with $\delta$ the declination and $b$ the galactic latitude).
This reference sample probes different environments, i.e., field, groups, and the Virgo cluster located at 16.5 Mpc \citep{Gavazzi_and_Boselli_1999,Mei_et_al_2007,Cantiello_et_al_2018}.
The distinction between ETGs and LTGs is based on the morphology of galaxies, i.e. the presence of spiral arms or prominent spiral arms \citep{Cappellari_et_al_2011}.  
As the focus of the ATLAS$^{3D}$ project, ETGs also have integral field unit (IFU) kinematic information available. This enables one to separate ETGs between fast rotators, that have regular rotation patterns, and slow rotators. Such a distinction is relevant as it is linked to the formation pathways of ETGs. Such kinematic information is missing for the LTG sample, but it is expected that LTGs are all fast rotators as suggested by their spiral pattern.

We used deep images from the 3.6-meter Canada-France-Hawaii Telescope (CFHT)  obtained with the wide-field optical imager MegaCam. MegaCam offers a pixel size of 0.187 arcsecond per pixel and a field of view of $1^{\circ}\times 1^{\circ}$. We utilised images originating from four surveys carried out at the CFHT: the Mass Assembly of early-Type GaLAxies with their fine Structures survey \citep[MATLAS\footnote{MATLAS, \href{http://obas-matlas.u-strasbg.fr}{http://obas-matlas.u-strasbg.fr}}, ][]{Duc_et_al_2015}; the Canada-France Imaging Survey \citep[CFIS\footnote{CFIS, \href{https://www.cfht.hawaii.edu/Science/CFIS/}{https://www.cfht.hawaii.edu/Science/CFIS/}}, ][]{Ibata_et_al_2017}; the Virgo Environmental Survey Tracing Ionised Gas Emission \citep[VESTIGE\footnote{VESTIGE, \href{https://mission.lam.fr/vestige/index.html}{https://mission.lam.fr/vestige/index.html}}, ][]{Boselli_et_al_2018}; and the Next Generation Virgo Cluster \citep[NGVS\footnote{NGVS, \href{http://astrowww.phys.uvic.ca/~lff/NGVS/Home}{http://astrowww.phys.uvic.ca/~lff/NGVS/Home}}, ][]{Ferrarese_et_al_2012}.

The main goals, observing strategies and available photometric bands of these four deep CFHT surveys differ. NGVS imaged the 104 square degrees of the Virgo cluster in four bands ($u$, $g$, $i$, $z$ and a few poitings in the $r$-band). VESTIGE, a blind narrow-band H$\alpha+$[NII] imaging survey, covered the same footprint as NGVS. VESTIGE also acquired $r$-band images of the whole cluster for calibration purposes and to subtract the continuum emission. 

MATLAS targeted nearby massive ETGs from the ATLAS$^{3D}$ sample through pointed observations in $g$ and $r$-bands (as well as in $u$ and $i$ for some galaxies).  
CFIS is a blind survey targeting 5,000 square degrees in the Northern Hemisphere in $u$ and $r$-bands, with additional bands available from observations made with other telescopes as part of the Ultraviolet Near Infrared Optical Northern Survey (UNIONS) project\footnote{The UNIONS project is a collaboration of wide field imaging surveys of the northern hemisphere. UNIONS consists of the Canada-France Imaging Survey (CFIS), conducted at the 3.6-meter CFHT on Maunakea, members of the Pan-STARRS team, and the Wide Imaging with Subaru HyperSuprime-Cam of the Euclid Sky (WISHES) team. CFHT/CFIS is obtaining $u$ and deep $r$ bands; Pan-STARRS is obtaining deep i and moderate-deep z band imaging, and Subaru/WISHES is obtaining deep z band imaging.}. All the NGVS bands, MATLAS bands, VESTIGE $r$-band and CFIS $r$-band images were processed through the LSB-optimised reduction pipeline Elixir-LSB \citep[Cuillandre, private communication,][]{Duc_et_al_2015} to enhance any faint structures. As a result of this processing, surface brightness limits of 29 mag$\,$arcsec$^{-2}$ in the $g$-band for NGVS \citep{Ferrarese_et_al_2012}, 28.9 mag$\,$arcsec$^{-2}$ for MATLAS $r$-band, 28.3 mag$\,$arcsec$^{-2}$ for CFIS $r$-band (Cuillandre, private communication), and 27.2 mag$\,$arcsec$^{-2}$ for VESTIGE $r$-band\footnote{\cite{Boselli_et_al_2018} reach 25.8 mag$\,$arcsec$^{-2}$ at 1$\sigma$ for a scale of 2.8 arcsec which can be converted to about 27.2 mag$\,$arcsec$^{-2}$ at 1$\sigma$ for a scale of 10 arcsec.} \citep{Boselli_et_al_2018} are reached. 
However, these values depend on the definition used, and can vary from one image to the other. Therefore, in Appendix \ref{section:appendix-sblim} we consistently re-compute the depths of our surveys using the $\mu_{lim, (3\sigma; 10 \arcsec\times 10 \arcsec)}$ surface brightness limit definition. We find a $r$-band $\mu_{r, lim (3\sigma; 10 \arcsec\times 10 \arcsec)}$ limit of 29.2 mag$\,$arcsec$^{-2}$ for MATLAS, 28.7 mag$\,$arcsec$^{-2}$ for CFIS and 29.1 mag$\,$arcsec$^{-2}$ for VESTIGE.  

The important aspect is that the same reduction pipeline was applied on images from the same telescope, instrument, and band\footnote{In Appendix \ref{section:appendix-filters_set} we discuss the impact of the filter change on MegaCam between CFIS/VESTIGE and MATLAS/NGVS}, which made the deep images look relatively similar.
The work in this study is based on the LSB $r$-band images from CFIS, MATLAS and VESTIGE, which was supplemented with $g$-band data for colour information, when available from NGVS and MATLAS. The $r$-band was chosen as the reference band as it is the only CFIS band on which the Elixir-LSB pipeline has been run; and because MATLAS and VESTIGE also had available Elixir-LSB-processed $r$-band images. Table \ref{table:data-surveys} summarises the properties of these CFHT surveys.

\begin{table*}
    \caption{Properties of the CFHT surveys utilised in this work.}
    \label{table:data-surveys}
   \centering
    \begin{threeparttable}  
    \footnotesize
    \begin{tabular}{ccccc}
    \hline
    Survey & Band used\tnote{a}  & Exposure time& Targets &Footprint area \\
    && [s] & & [deg$^2$] \\
    \hline
    NGVS\tnote{b}    & \textit{Old} $g$ & $5 \times 634 $& Virgo cluster & 116\\
    VESTIGE & \textit{New} $r$ & $12 \times 60$   & Virgo cluster & 116 \\
    MATLAS  & \textit{Old} $r$ (+ \textit{old} $g$)  & $7 \times 345$  & Group ETGs & 310\\
    CFIS    & \textit{New} $r$ & Varying \tnote{c} & Blind survey & 5,000 \\
    \hline
    \end{tabular}
    \begin{tablenotes}
        \item[a] MegaCam's filter set has been changed in 2015; the filters used before 2015 are denoted as \textit{Old} and the ones in use afterwards are denoted as \textit{New}. See Appendix \ref{section:appendix-filters_set} for more details.
        \item[b] NGVS $g$-band images were only used for colour information, not for the detection and analysis of tidal features.
        \item[c] The integration time is adjusted based on instant observing conditions (atmosphere, image quality, sky background) to ensure that the full depth within the 3 exposures will be uniform across the whole survey footprint, \href{https://www.cfht.hawaii.edu/Science/CFIS/cfissurvey.html}{https://www.cfht.hawaii.edu/Science/CFIS/cfissurvey.html}.
    \end{tablenotes}
\end{threeparttable}
\end{table*}

We used Multi-Order Coverage maps \citep[MOCs,][]{Fernique_et_al_2014} to determine which galaxies from the ATLAS$^{3D}$ reference sample had available CFHT LSB-optimised $r$-band images. We obtained a final sample of 475 galaxies, among which are 244 ETGs and 231 LTGs. Table \ref{table:data-nb_gal_survey} details the number of galaxies as a function of the survey and environment. The complete list of our galaxies and  their basic properties are presented in Table \ref{table:appendix-table_properties_galaxies}. Some galaxies present in CFIS images had already been observed by MATLAS. In that case, we preferentially used the deeper MATLAS images.

For data processing, we followed the procedures detailed in \cite{Sola_et_al_2022}. We cropped the  images around the galaxies of interest, keeping a field of view of $31\arcmin\times31\arcmin$, large enough to cover the extended LSB structures around them. This corresponds to a field of view of about 45 kpc $\times$ 45 kpc at a distance of 5 Mpc; 150 kpc $\times$ 150 kpc at 16.5 Mpc; and 415 kpc  $\times$ 415 kpc at 46 Mpc. Images were then binned by a factor of three, and a modified inverse hyperbolic sine scaling, asinh, was applied to enhance the faintest structures. The surface brightness (SB) maps in units of mag$\,$arcsec$^{-2}$ were computed using Equation 2 of \cite{Sola_et_al_2022}. Finally, the images were converted to the HiPS format \citep[see][]{hips} and uploaded to the \textit{Jafar} annotation tool.
In addition to our deep images, we used the shallower composite colour images of the DR1 PanSTARRS\footnote{PanSTARRS, \href{https://panstarrs.stsci.edu/}{https://panstarrs.stsci.edu/}} imaging survey \citep{Chambers_et_al_2016}, available at CDS,   to characterise the brightest inner regions of the galaxies.  PanSTARRS covers the entire sky north of $- 30$ $^{\circ}$ in declination in five bands ($g$, $r$, $i$, $z$ and $y$). In Figures \ref{fig:appendix-thumbnails_composite_ETGs} and \ref{fig:appendix-thumbnails_composite_LTGs}, we present color-composite images of our galaxies, with the greyscale corresponding to the deep image and the colour image to a combination of PanSTARRS $r,g,i$ images.

Our final sample of 475 galaxies is composed of 94\% of the 260 ETGs and 38\% of the 611 LTGs from the ATLAS$^{3D}$ reference sample. They are located in different environments, from the field to a cluster. Note that we only have one cluster environment, Virgo, which is still dynamically young and not representative of all the cluster types \citep[e.g.,][]{Binggeli_et_al_1987,Boselli_et_al_2014}. We parametrise the environment by the local environment density  $\rho_{10}$, defined as the volume density in Mpc$^{-3}$ of galaxies inside a sphere centred on the target galaxy and including the ten nearest neighbours \citep{Cappellari_et_al_2011b}. This mean number density is based on the adaptative method of \cite{Dressler_1980}.
Galaxies in the Virgo cluster typically have $\log_{10}(\rho_{10}) > -0.4$. 
Our sample includes 98\% (79\%) of the Virgo ETGs (LTGs) found in the ATLAS$^{3D}$ reference sample. Outside the Virgo cluster, our sample contains 93\% (33\%) of the field ETGs (LTGs) in the ATLAS$^{3D}$ reference sample.

\begin{figure*}
  \centering
  \includegraphics[width=0.45\linewidth]{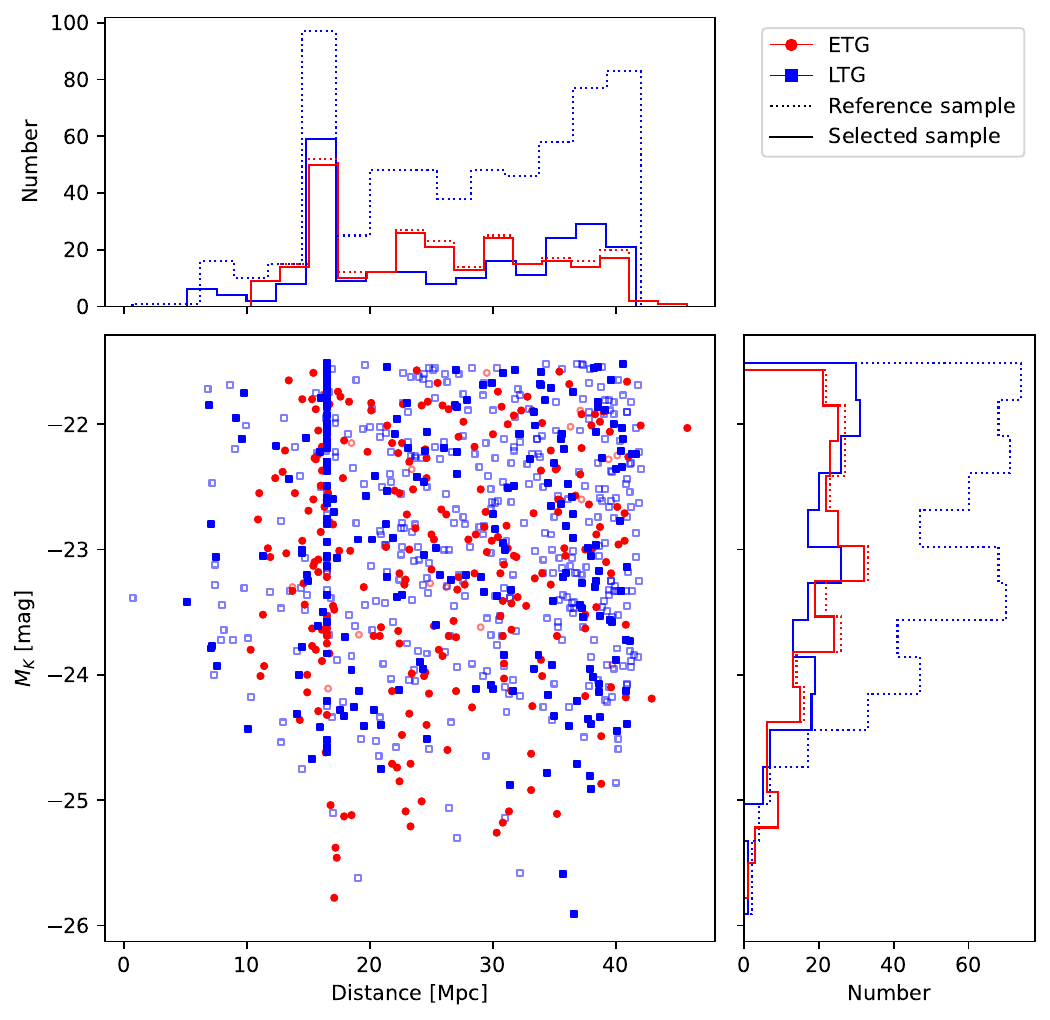}
  \hspace{0.5cm}
   \includegraphics[width=0.45\linewidth]{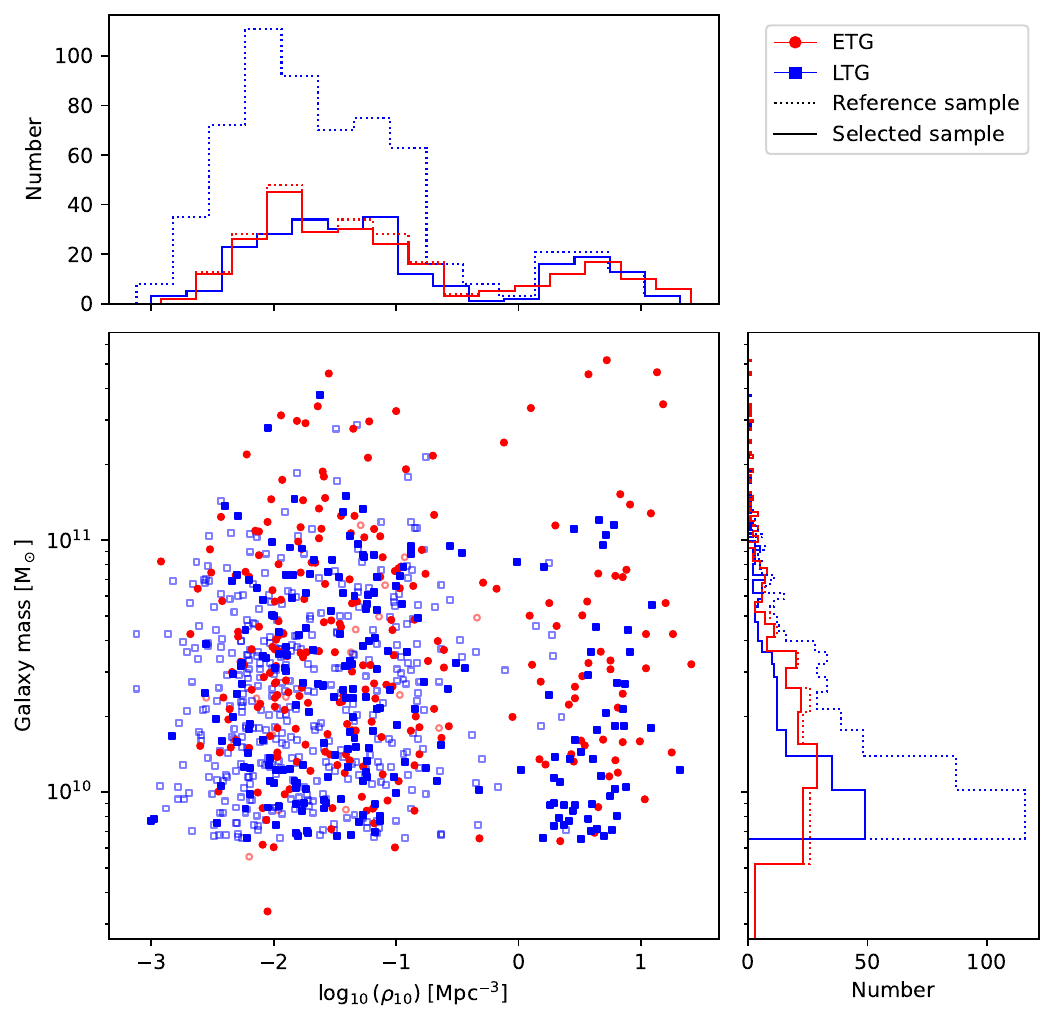}    
   \caption{Main properties of our final sample of 475 galaxies, compared to the reference sample from the ATLAS$^{3D}$ project. ETGs are plotted in red circles and LTGs in blue squares. The solid lines and filled circles correspond to our final sample, while the dotted lines and open circles correspond to the reference sample. \textit{Left:} Scatter plot of the absolute $K$-band magnitude M$_K$ (in mag) versus the distance (in Mpc) of galaxies from our final sample, the ATLAS$^{3D}$ reference sample, and their marginal histograms.  
   \textit{Right:} Scatter plot of the galaxy stellar mass (in M$_\odot$) as a function of the environment density $\rho_{10}$ (in Mpc$^{-3}$) of galaxies from our final sample, the ATLAS$^{3D}$ reference sample, and their marginal histograms. Galaxies in the Virgo cluster typically have $\log_{10}(\rho_{10}) > -0.4$.}
    \label{fig:data-properties_sample}
\end{figure*}

In Figure \ref{fig:data-properties_sample}, we present the main properties of our 475 galaxies. The left panel shows the absolute $K$-band magnitude, M$_K$, as a function of the distance, $D$, while the right panel displays the stellar mass as a function of the environment density, $\rho_{10}$.
The overdensity of galaxies around 16.5 Mpc is due to the Virgo cluster. 
We do not have in our sample some of the closest and furthest LTGs from the ATLAS$^{3D}$ reference sample; as those reference LTGs are located at a median distance of 29 Mpc against 25.4 Mpc for our selected LTGs.
As seen in the right plot, our selected ETGs are more massive (median mass of $3.3 \times 10^{10} M_\odot$) than our selected LTGs (median mass of $2.2 \times 10^{10} M_\odot$), which is also the case for the ATLAS$^{3D}$ reference sample and for the nearby Universe, as seen for instance by the S$^4$G catalogue \citep[][]{Sheth_et_al_2010,Munoz-Matos_et_al_2015} further extended and updated by \cite{Watkins_et_al_2022,Sanchez-Alarcon_et_al_2025}. The marginal histogram over environment density shows two trends. First, our selected ETGs and LTGs are located in similar environments, which makes further comparisons simpler. Second, compared to the reference LTGs, we miss some LTGs in regions of low- and intermediate-densities (i.e., in the densest regions of galaxy groups, below the Virgo cluster density). We are indeed  biased towards LTGs located in the ETG-rich groups observed by MATLAS\footnote{MATLAS pointings targeted group ETGs, but in some images group LTGs were also observed.}, which is not fully compensated by the availability of the blind CFIS images. 

\begin{table*}
\caption{Number of galaxies studied in this work, according to the  survey used to image them, their morphological type and Virgo membership. 
}             
\label{table:data-nb_gal_survey}     
\centering                         
\begin{tabular}{c c c c c |c}        
\hline              
Type & Virgo ETGs & Virgo LTGs & Field ETGs & Field LTGs & Total \\   
\hline                        
    MATLAS & 5 & 4 & 174 & 49 & 232 \\
    CFIS & 0 & 0 & 7 & 113  & 120  \\
    VESTIGE & 52 & 51 & 6 & 14 & 123 \\
\hline    
    Total & 57 & 55 & 187 & 176 & 475 \\
\hline                                  
\end{tabular}
\end{table*}

\section{Methods} \label{section:methods}
In this section, we present \textit{Jafar}, the annotation tool used to study the LSB tidal features in our deep images. We describe how we characterised these structures, in terms of shape and luminosity.

\subsection{The \textit{Jafar} annotation server}\label{section:jafar}
\cite{Sola_et_al_2022} introduced a web-based annotation tool that enables contributors, preferentially trained experts, to delineate precisely the shapes of LSB structures superimposed on deep images. Here we name it \textit{Jafar}\footnote{\textit{Jafar}, \href{https://jafar.astro.unistra.fr}{https://jafar.astro.unistra.fr/}}, standing for Just Annotate Features for Astronomical Results\footnote{And a nod to the names of the services provided by CDS, Aladin, Simbad or Vizier.}. This user-friendly interactive tool provides efficient ways to visualise and navigate through images, draw features and label the annotations. 

Each galaxy can be annotated by several contributors\footnote{Contributors can see only their own annotations.}. Contributors are asked to delineate the contours of LSB features by choosing the most appropriate drawing shape among several (e.g. ellipses, polygons, circles, rectangles, or cubic Bézier curved lines) and attach a corresponding label. All drawn shapes are summarised in a table and can later be modified. Once the annotation is done, the coordinates in right ascension and declination of the contours of the shapes are stored in a database, along with the corresponding label. As contributors are asked to annotate every relevant feature, several drawn shapes are superimposed on each image.

Features of interest can be labeled as \textit{Inner galaxy}, \textit{Halo}, \textit{Tidal Tails}, \textit{Streams}, \textit{Shells}, \textit{Companion Galaxy}, \textit{Cirrus}, \textit{Ghost Reflections}, \textit{Satellite Trails}, \textit{Instrument} or \textit{High Background}. They can be categorized as one of two broad types: stellar light and contamination sources, the light of which hinders the annotation process.
Starting with stellar components, the \textit{Inner galaxy} is defined as the central region of the target galaxy, encompassing the most luminous features. It corresponds to the extent of the whole galaxy in the shallow PanSTARRS images. Our so-called \textit{Halo} corresponds to the extent of the galaxy in deep images, i.e., it encompasses the LSB outskirts. These subjective by-eye definitions enable us to quickly retrieve the outer shape of galaxies and their total luminosity. A proper study of the stellar haloes will be the focus of a follow-up paper, along with haloes' radial profiles, potential disc breaks and PSF-deconvolution. In Figures \ref{fig:appendix-thumbnails_composite_ETGs} and \ref{fig:appendix-thumbnails_composite_LTGs}, the \textit{Inner Galaxy} corresponds to the region in true colour, while the \textit{Halo} is the greyscale part.

For delineating tidal features, we follow the definitions established in the previous papers of this series \citep[e.g.,][]{Duc_et_al_2015,Bilek_et_al_2020, Sola_et_al_2022}. We refer to Figures 3 and 4 of \cite{Sola_et_al_2022} for illustrations of each type of structure defined here. We consider \textit{Tidal Tails} as elongated  material that appears to originate from the target galaxy and presumably formed during major mergers \citep[e.g.,][]{Arp_1966,Toomre_and_Toomre_1972,Mihos_1995}, although they could also form from non-merging flybys \citep[e.g.,][]{Duc_and_Renaud_2011}. On the contrary, \textit{Streams} likely originate from minor mergers \citep[e.g.,][]{Bullock_and_Johnston_2005, Belokurov_et_al_2006, Martinez_Delgado_et_al_2010}. The low-mass companions from which \textit{Streams} emanate may be visible on the image or not. Therefore orphan structures with no obvious progenitors are considered as streams. Although tails and streams are generated by the same process, i.e. tidal interaction with another companion galaxy, the distinction we make is based on the point of view of the target, the massive galaxy in the field. This enables us to obtain a more subtle disentanglement of the types of involved mergers (major or minor)\footnote{In the absence of a ground truth and from photometric data only, disentangling streams from tails in ambiguous cases might be complicated. Although \cite{Mancillas_et_al_2019} showed in simulated images that the detectability of tails and streams does not depend on projection effects, some subjectivity in the labelling remains, which is why we combine both tidal features types in some analyses.}. 

\textit{Shells} are arc-shaped, often concentric, features that tend to form during intermediate-mass radial mergers \citep[e.g.,][]{Prieur_1990, Ebrova_2013, Duc_et_al_2015, Pop_et_al_2018}. Nearby \textit{Companion} galaxies with a mass similar to the target galaxy and a slight velocity difference (about 200 km$\,$s$^{-1}$)\footnote{This threshold was chosen to ensure that the two galaxies currently involved in a tidal interaction are able to produce visible tidal debris. The velocity information is taken from the Simbad database, as \textit{Jafar} (relying on Aladin) enables to overlay catalogues on the images during the annotation process.} are also delineated. 

Contributors also annotated several contaminants, including \textit{Cirrus}, the dust clouds from our own Galaxy that scatter the optical light, and  circular \textit{Ghost reflections}, i.e  artificial haloes around bright sources, such as foreground stars or  galaxies that result from internal light reflections in the instrument. Moreover \textit{Satellite Trails},  \textit{Instrument} artefacts such as CCD gaps and regions having  \textit{High Background} values of unknown origin  may also be drawn with the annotation tool. We used the \textit{High Background} label in this study when the LSB structures of interest were contaminated.

\cite{Sola_et_al_2022} addressed the issue of determining which annotation to keep when a given structure was annotated by several contributors to avoid duplicates. To that end, we applied in this paper the selection process described in Section 4.8 of \cite{Sola_et_al_2022}, and we briefly summarise it in Section \ref{section:results-census_features}. We only used annotations conserved after that selection process (unless explicitly mentioned otherwise).

All our 475 galaxies have been annotated by at least two contributors out of four. One contributor annotated 100\% of the galaxies, while three others annotated 74\%, 69\% and 22\% of our sample. 
Table \ref{table:results-nb_annotations} summarises the number of annotations stored in the database as a function of their type. The middle column indicates the total number of annotations by all contributors, including the duplicates, while the right column indicates the number of annotations kept after the selection process. We eliminated duplicates, merging common annotations. We only kept tidal tails whose progenitor is the target galaxy in order to avoid counting tails originating from a companion galaxy twice. Twenty haloes were impossible to annotate due to contamination sources (bright foreground star or high background) overlapping with the galaxy. These 199 tidal tails and 100 streams are used to derive the results presented in this paper. 

We publicly release Tables \ref{table:appendix-nb_features} and Table \ref{table:appendix-list_features}, which list all the individual tails and streams along with some of their properties such as geometry, SB and luminosity, as well as the number of tidal features per galaxy. In Appendix \ref{section:appendix-coordinates_features}, we provide a file containing the coordinates of the contours of the individual features (presented as polygons) in a format readable by Aladin and SAOImageDS9.
Our database of LSB features is also of special utility for training automated ML-based algorithms. For instance, \cite{Richards_et_al_2024,Richards_et_al_2024b} trained a deep learning algorithm on our annotation database to automatically segment galactic structures and cirrus in deep images.
\begin{table}
\caption{Number of annotations stored in the database as a function of their type. \textit{Total} refers to the total number of annotations made by all users, while \textit{Selected} reflects the number of annotations kept after the selection process, that affects \textit{Inner Galaxy}, \textit{Halo}, \textit{Tidal Tails} and \textit{Streams}, and used in our analysis.}

\centering  
\label{table:results-nb_annotations}  
\begin{tabular}{ccc} 
    \hline 
    Annotation type & Total  & Selected\\
    \hline
    Inner Galaxy    &  1259 & 475 \\
    Halo            & 1193 & 455  \\
    Tidal Tails     & 616  & 199  \\ 
    Streams         & 202  & 100 \\
    Shells          & 311  & 311   \\ 
    Companion Galaxy& 991  & 991 \\ 
    High Background & 1307 & 1307 \\
    Ghosted Halo    & 5343 & 5343 \\
    Cirrus          & 317  & 317 \\
    Satellite Trail & 114  & 114 \\
    Instrument      & 467  & 467 \\
    Total           & 11832 & 10081  \\ 
    \hline
\end{tabular}
\end{table}

\subsection{Computing the shapes and luminosity of LSB features} \label{section:data-geometry_flux_SB}
The database of drawn annotations provides quantitative information about LSB structures. Geometrical analyses are presented in \cite{Sola_et_al_2022}. To summarise, we computed geometrical properties such as the area, length, width and radii of shells from the contours of the annotations. We also created masks of the annotations that we applied on the surface brightness maps to automatically retrieve characteristic SB values of the annotated features. A similar process yielded $g-r$ colour values. 

In this work, we refine our SB measurements by performing background subtraction to remove the excess light coming from bright sources or from overlapping annotations. Previously, we used the median value as the most representative SB value in each annotation, which enabled us to obtain results very quickly from our masks. Here, we developed a method to obtain more precise measurements of the luminosity and SB inside annotated features at the cost of longer computation times. The process is described in Appendix \ref{section:appendix_flux_estimation}. Briefly, even though the background is supposed to be constant after being processed by Elixir-LSB, local variations still exist due to contamination sources and must be removed. To that end, we remove bright sources from the masks of the annotations, we then estimate a new background value that we later subtract from the masks. These background-corrected values of luminosity and SB are presented in this paper.

\subsection{Statistical tests}
In the following sections,  tests are applied to determine whether trends are statistically significant. The t-test\footnote{\href{https://docs.scipy.org/doc/scipy/reference/generated/scipy.stats.ttest_ind.html}{Scipy's stats.ttest\_ind}} \citep{Student_1908} tests the null hypothesis that two independent samples have the same mean, while the 2-sample Kolmogorov-Smirnov (KS)\footnote{\href{https://docs.scipy.org/doc/scipy/reference/generated/scipy.stats.ks_2samp.html}{Scipy's 2-sample KS test}} \citep{Hodges_1958} tests the null hypothesis that two samples come from the same distribution. Mood's test\footnote{\href{https://docs.scipy.org/doc/scipy/reference/generated/scipy.stats.median_test.html}{Scipy's Mood's median test}} \citep{Mood_1950} is used to test the null hypothesis that two samples come from populations with the same median. The Z-test, testing can be applied to check whether the proportions from two populations are the same. Pearson's r\footnote{\href{https://docs.scipy.org/doc/scipy/reference/generated/scipy.stats.pearsonr.html}{Scipy's Pearson's r test}} \citep{Pearson_1895} and Spearman's r\footnote{\href{https://docs.scipy.org/doc/scipy/reference/generated/scipy.stats.spearmanr.html}{Scipy's Spearman's r test}} tests \citep{Spearman_1904} are applied to test linear correlations and monotonic relationships, respectively. If the conditions of applications of the tests are verified, and if the p-value of the tests is lower than a given threshold, namely 0.05 for a confidence level of 5\% (or two-sigma), the null hypothesis can be rejected. We chose a threshold of 5\% to reject the null hypothesis and to consider the result of a test to be statistically significant.

\section{Global statistical results}\label{section:results-properties_features}

\subsection{Incidence of tidal features}\label{section:results-census_features}
The total fraction of galaxies in our sample showing any sign of interaction (i.e. one or more features of any type) is 36\%. If we distinguish between the types of features, 23\% of galaxies host tidal tails, 15\% host streams and 12\% host shells. We also counted the number of tidal features around each galaxy. To that end, we used the selection process described in Section 4.8 of \cite{Sola_et_al_2022} for tidal tails and streams. Briefly, for tails and streams identified around each galaxy, we group annotations from multiple contributors by assessing the percentage of overlap between features. We retain only the annotation with the largest area, as more experienced contributors typically detect fainter and thus more extended structures \citep[e.g.,][]{Bilek_et_al_2020}.
A different process is needed for shells. Indeed, as shells are delineated using single-pixel-wide lines, applying a selection criterion based on zero intersection would not be meaningful. 
In addition, we are facing the issue of the precise delineation of shells, which depends on the expertise of the contributors. If a shell is ambiguous and not so bright, it can be drawn as a single long arc by an expert contributor, while it can be cut in several parts by a less-expert contributor. Hence we consider the mean number of shells per contributor and galaxy for the census of shells\footnote{We assume here that our four contributors have similar levels of expertise.}. The mean number of tidal features per galaxy, detailed by survey and morphological type, can be found in Table \ref{table:mean_nb_debris_per_gal}. 

\begin{table}
\caption{Mean number of tidal features per galaxy, taking into account all the galaxies, even those without tidal features. In parenthesis is indicated the mean number of features per galaxy computed only for the galaxies that do present tidal debris. }         
\label{table:mean_nb_debris_per_gal}     
\centering                         
\begin{tabular}{c c c c}        
\hline              
Type & Tidal Tails & Streams & Shells  \\   
\hline                        
    MATLAS ETGs & 0.28 (1.89)  & 0.28 (1.42) & 0.44 (2.82) \\
    CFIS ETGs   & 0.00 (--)     & 0.29 (1.0)  & 0.00 (--) \\ 
    VESTIGE ETGs& 0.38 (1.47)  & 0.16 (1.12) & 0.30 (2.5) \\
    Total ETGs  & 0.30 (1.74)  & 0.25 (1.35) & 0.40 (2.76)\\
&&&\\
    MATLAS LTGs & 0.60 (2.0)   & 0.06 (1.0)  & 0.22 (1.48) \\ 
    CFIS LTGs   & 0.50 (1.87)  & 0.27 (1.43) & 0.14 (1.71) \\ 
    VESTIGE LTGs& 0.58 (1.65)  & 0.08 (1.67) & 0.13 (1.42) \\
    Total LTGs  & 0.55 (1.83)  & 0.16 (1.41) & 0.15 (1.55) \\
&&&\\ 
    Total       & 0.42 (1.79)  & 0.21 (1.37) & 0.28 (2.28) \\
\hline                                  
\end{tabular}
\end{table}

Tidal tails are more numerous around LTGs than ETGs, with a mean number of 0.55 and 0.30 per galaxy, respectively, which is statistically significant according to the 2-sample KS-test (p-value $=$ 0.04 so we can reject the null hypothesis that the samples come from the same distribution). There is no significant difference in the number of streams for ETGs and LTGs, in terms of means (p-value $=$ 0.08 for the t-test) and distributions (p-value $=$ 0.54 for the 2-sample KS-test). Although shells seem more numerous around ETGs than LTGs, this is not significant as the 2-sample KS-test revealed similar distributions (p-value of 0.46). The conditions of application of the t-test were not verified for the number of tails and streams.

The higher fraction of tails around LTGs than ETGs is mainly explained by contamination from ongoing tidal interactions, which is higher for LTGs than ETGs. Indeed, disk galaxies are more susceptible to producing distinct prominent antennae-like features, due to their low local velocity dispersion and thinness, while elliptical galaxies will trigger more diffuse plume-like tails \citep[e.g.,][]{Duc_and_Renaud_2011}. A similar fraction of streams observed around ETGs and LTGs was expected. Indeed, as streams originate from lower-mass companions, the morphological type of the primary galaxy should not directly impact the companions. For shells, the important parameters for their formation are the impact parameters, and not so much the morphological type of the primary galaxy.

\subsection{Luminosities of LSB features}\label{section:results-luminosities}
The luminosity of tidal features, derived from the observed flux, is an important quantity to measure as it can be compared with that of simulated features. The flux was automatically retrieved through aperture photometry with the masks of the annotations, as described in Section \ref{section:data-geometry_flux_SB}. 

We calculated the fraction of the total flux emitted by the tidal features in each system. For each galaxy, we summed the flux in all the tidal features (tails and streams) and we divide it by the flux of the whole galaxy. This total galaxy flux includes both the inner regions and the extended LSB stellar outskirts but explicitly excludes tidal features. We note that in some cases there might be some overlap between a part of the tidal features and the outskirts \textit{Halo} annotation, so we may be slightly overestimating the tidal feature luminosity fraction. An accurate photometric estimate would require a proper disentanglement between tidal features and stellar halo in these ambiguous regions, with in-depth analysis and precise galaxy modelling.

\begin{figure*}
    \centering
    \includegraphics[width=0.45\linewidth]{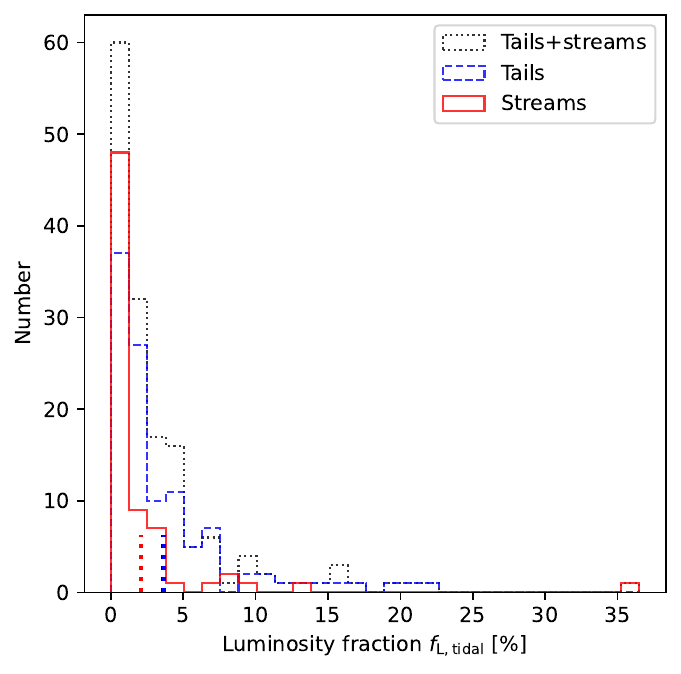}
    \hspace{0.1cm}  
    \includegraphics[width=0.45\linewidth]{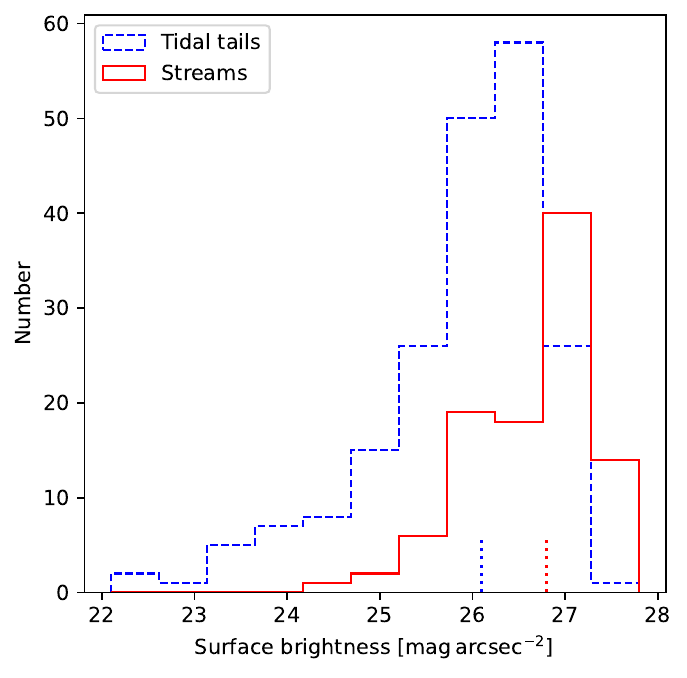}
    \caption{Histograms of the luminosity fractions f$_{\rm L,tidal}$ and median surface brightness of tidal tails and streams (presented only for galaxies that do have tidal features). \textit{Left}: f$_{\rm L,tidal}$ of tails (blue), streams (red) and tails and streams combined (black). The vertical dotted lines represent the mean values of the respective distributions. \textit{Right}: median surface brightness of tails (blue) and streams (red).
    The vertical dotted lines represent the median values of the respective distributions.}
    \label{fig:results-histogram_luminosityfraction}
\end{figure*}

For the galaxies that do show tidal features, we present the tidal feature luminosity fraction f$_{\rm L,tidal}$ in the left panel of Figure \ref{fig:results-histogram_luminosityfraction}.
Tails and streams account for only a few percent of the total galaxy luminosity. The distributions of the luminosity fractions are dominated by low values, with a mean (median) fraction of 3.6\% (1.8\%) for tails and streams combined. For tails only, the mean (median) luminosity fraction is 3.7\% (1.9\%) and it is 2.1\% (1.8\%) for streams only.

We present the median surface brightness in tails and streams in the right panel of Figure \ref{fig:results-histogram_luminosityfraction}. Streams are fainter than tails, with a median surface brightness of 26.8 and 26.1 mag$\,$arcsec$^{-2}$, respectively, with Mood's test confirming this difference to be statistically significant. The faintest stream reaches a median SB of 27.8 mag$\,$arcsec$^{-2}$, about one magnitude brighter than our SB limit.

\section{Trends as a function of the mass of the host galaxy} \label{section:results-effect_galaxy_mass}
In this section, we investigate the relationships between the incidence and luminosity of tidal features, and the host galaxy's mass. The computation of the stellar mass is presented in Appendix \ref{section:appendix-galaxy_sample}.
As galaxies are spread over a wide range of masses (see Figure \ref{fig:data-properties_sample}), it is necessary to group data into bins of mass containing approximately the same number of galaxies, i.e. different bin widths. In each bin, the mean or median values of the quantity of interest are computed, and represented on a scatter plot. 

\begin{figure*}
  \centering
  \includegraphics[width=0.5\linewidth]{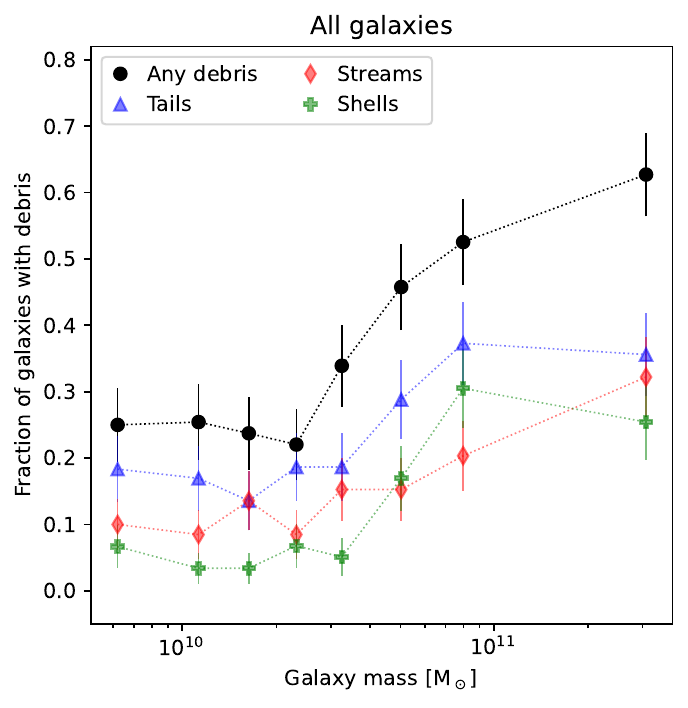}
  \hspace{5cm}
  \vspace{0.1cm}
   \includegraphics[width=0.4\linewidth]{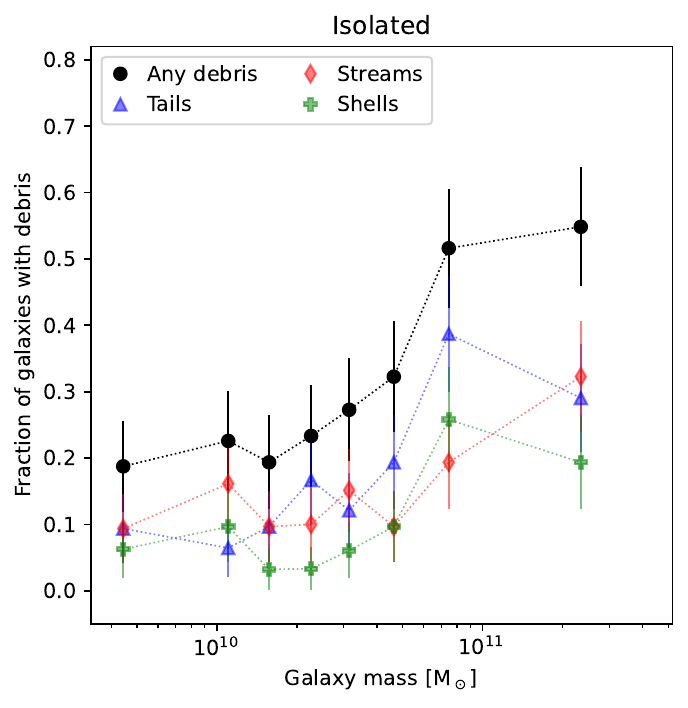}    
    \hspace{0.1cm}
   \includegraphics[width=0.4\linewidth]{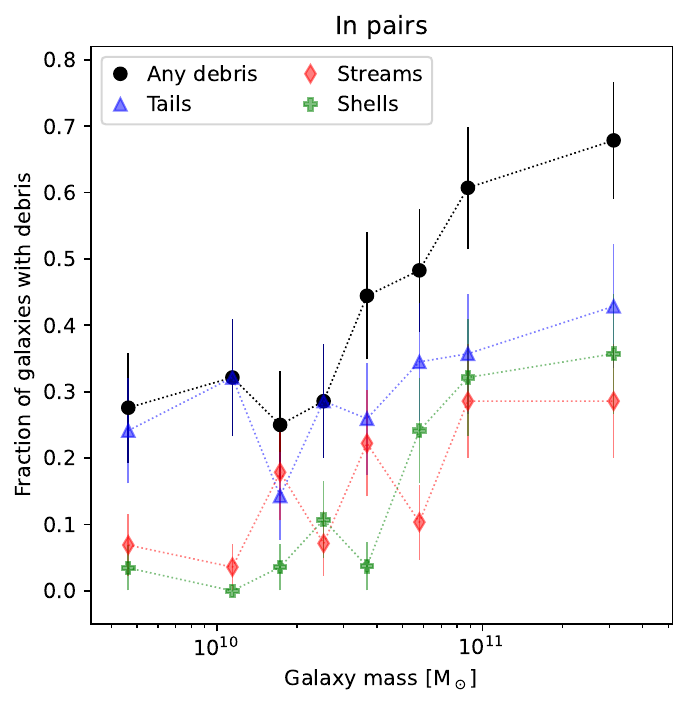}    
   \caption{Fraction of galaxies that have tidal features as a function of the galaxy stellar mass (in M$_\odot$), per mass bin. Each bin  contains approximately the same number of galaxies (i.e. about 60 galaxies per bin for the top panel, about 31 for the bottom left and 28 for the bottom right). The fraction of galaxies hosting any type of debris is plotted in black, galaxies hosting tails in blue, streams in red and shells in green. The error bars represent the standard deviation on proportions in each bin. \textit{Top}: Fraction of galaxies with debris considering all galaxies together. \textit{Bottom left}: Fraction of galaxies with debris only considering isolated (i.e. post-merger) galaxies (see Section \ref{section:results-effect_environment}). \textit{Bottom right}: Fraction of galaxies with debris only considering  galaxies in pairs, i.e. that can be undergoing tidal interactions with a massive companion.}
    \label{fig:results-fraction_galdebris_vs_mass}
\end{figure*}

\begin{figure*}
  \centering
  \includegraphics[width=0.4\linewidth]{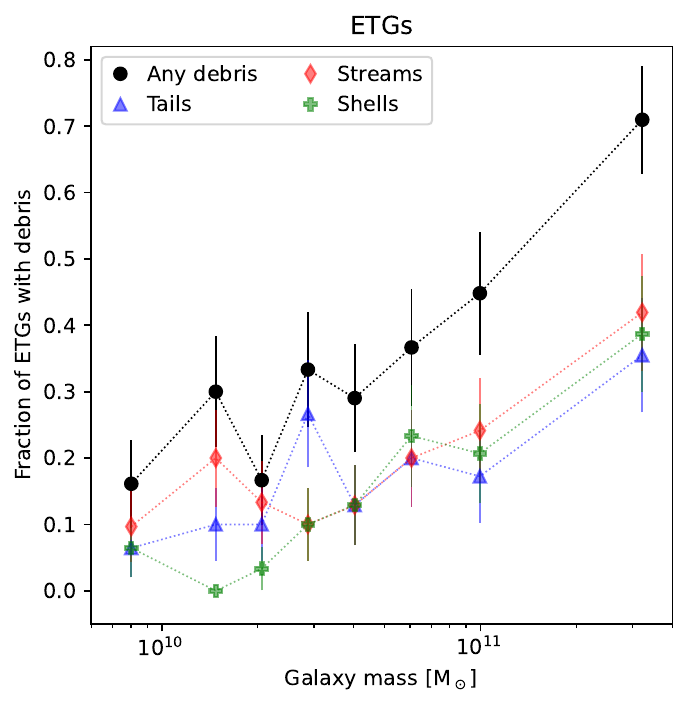}
\hspace{0.1cm}
   \includegraphics[width=0.4\linewidth]{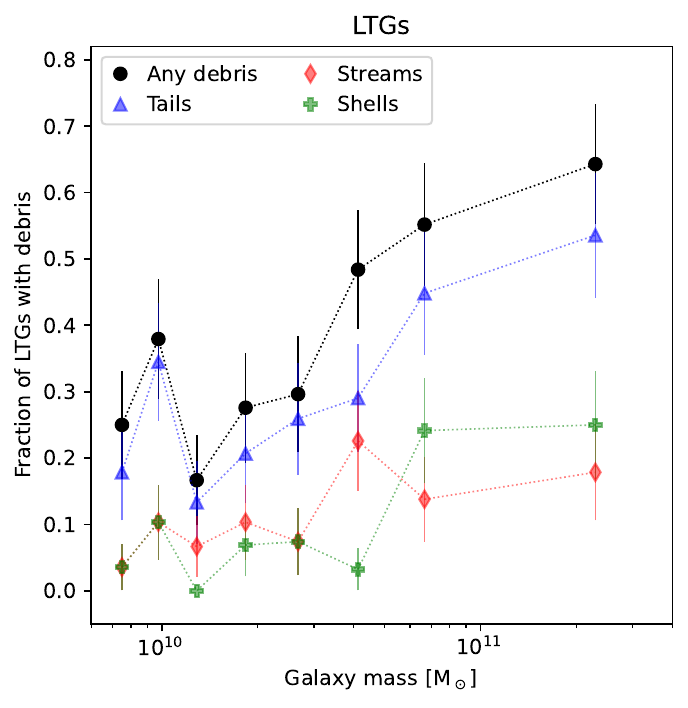}    
   \caption{Fraction of ETGs (\textit{left}) and LTGs (\textit{right}) that have tidal features as a function of the galaxy stellar mass (in M$_\odot$), per mass bin. Each bin contains approximately the same number of galaxies, i.e. about 30 galaxies per bin in the left plot and 29 in the right plot. The fraction of galaxies hosting any type of debris is plotted in black, galaxies hosting tails, in blue, streams in red and shells in green. The error bars represent the standard deviation on proportions in each bin.}
    \label{fig:results-fraction_galdebris_ETGs_LTGs_vs_mass}
\end{figure*}

In the top panel of Figure \ref{fig:results-fraction_galdebris_vs_mass}, we study the evolution of the fraction of galaxies hosting tidal debris as a function of galaxy mass and show the results\footnote{The bottom panel of Figure \ref{fig:results-fraction_galdebris_vs_mass} presents results that will be studied in Section \ref{section:results-effect_environment}, where we will separate features from ongoing interactions in pairs of galaxies (bottom right) and features around isolated galaxies (or `post-merger') (bottom left). However, we consider in this section all galaxies together, whether interacting or isolated.}. 
The fraction of galaxies hosting debris increases with galaxy mass for all types of tidal features. The fraction of galaxies with any type of debris is around 20-30\% for galaxy masses lower than $3\times10^{10}M_\odot$, but it increases sharply up to 60\% for the highest mass galaxies. This increase is statistically significant at a 5\% confidence level, as Spearman's test revealed a positive monotonic relation between both quantities (p-value of 0.02). This trend is driven by all three types of tidal features, as the fraction of galaxies with tails goes from approximately 20\% to 30\%, from 5\% to 22\% for shells and from 12\% to 25\% for streams, with a sharper increase around the same mass threshold.
When we separate the galaxies as a function of the morphological type, a similar global increase in mass is observed (see Figure \ref{fig:results-fraction_galdebris_ETGs_LTGs_vs_mass}). Tails primarily influence the trend for LTGs, whereas the steeper rise observed for ETGs is attributed to all types of tidal features.

To study the evolution of the luminosity fraction of tails and streams (f$_{\rm L,tidal}$) as a function of galaxy mass, we divided our sample between low- and high-mass galaxies based on a mass threshold of $M>4\times 10^{10} M_\odot$. We also confirmed that the results from the statistical tests do not change if we take mass thresholds of $3,5,6,7$ or $8\times 10^{10} M_\odot$. We compared f$_{\rm L,tidal}$ in the low- versus high-mass samples. The t-test revealed similar means (p-value of 0.068), but the distributions and the median are statistically significantly different from the 2-sample KS-test and Mood's test. Hence, more  massive galaxies have slightly more luminous tidal features. This result will be discussed in Section \ref{section: discussion-mass-env-combined}, in particular with respect to surface brightness limit and tidal feature detectability.

\section{Trends as a function of the environment}\label{section:results-effect_environment}
In this section, we investigate the dependence between the environment and collisional debris. The environment can be parametrised in different ways. We distinguish between the small-scale environment (presence of a nearby companion susceptible to generating tidal interactions) and the larger-scale one (cluster, field, group). We caution that our so-called large-scale environment is based on the local environment density $\rho_{10}$ distinguishing the Virgo cluster from the group/field, but it is not related to the cosmic web structures like filaments, clusters and voids.

In the following plots, galaxies are grouped in bins of environment density $\rho_{10}$ that contain approximately the same number of galaxies per bin.
The small-scale environment was defined using the annotations made using \textit{Jafar}. Contributors were asked to delineate a companion galaxy as a massive nearby galaxy if it was considered close enough to be able to trigger detectable tidal interactions. We set a threshold of a projected distance of 100 kpc\footnote{This arbitrary threshold is low enough to ensure that any tidal interaction would trigger detectable tidal features.}. Note that we were not able to check the actual mass of the galaxy (except if it was a galaxy from the reference sample of ATLAS$^{3D}$); we estimated the galaxy as massive if its size was close to the one of the target galaxy, i.e. not a dwarf or a much smaller companion. In addition, galaxies can be close in projection but their actual distance can be larger, and we do not correct for this effect.

Galaxies with no such companion are referred to as `isolated’ or `post-merger', while galaxies having at least one companion are termed `in pairs’. For clarity, even if the galaxy has several companions, it is referred to as in pairs. With this definition, a galaxy in the Virgo cluster can be considered isolated, and a galaxy in the field/group can be in a pair. Separating on-going collisions (47\% of the sample) and isolated galaxies is highly relevant for our study. Indeed, while the young tidal features produced in on-going interactions give information about the future of the host galaxy, and a possible imminent merger, only the features from ancient mergers are able to trace the past mass assembly we wish to reconstruct. 

\begin{figure*}
  \centering
  \includegraphics[width=0.5\linewidth]{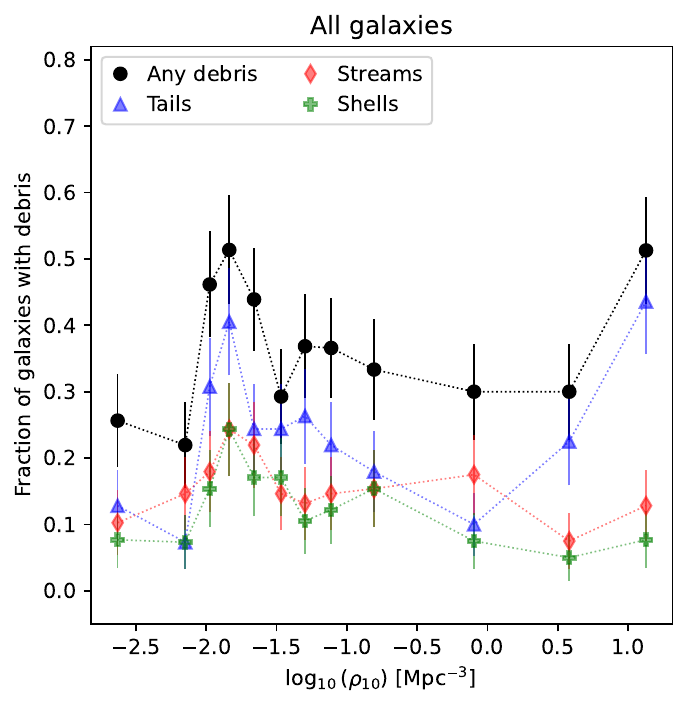}
  \hspace{2cm}
  \vspace{0.1cm}
   \includegraphics[width=0.4\linewidth]{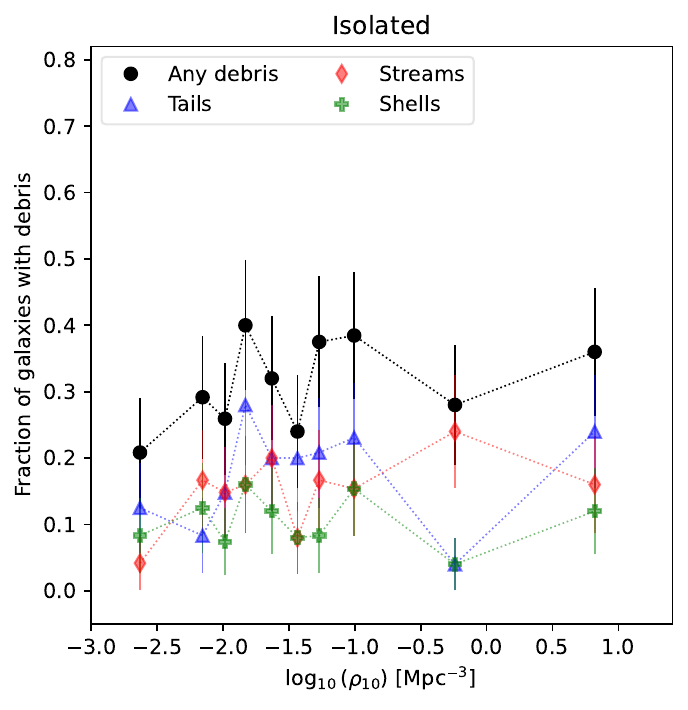}    
    \hspace{0.1cm}
   \includegraphics[width=0.4\linewidth]{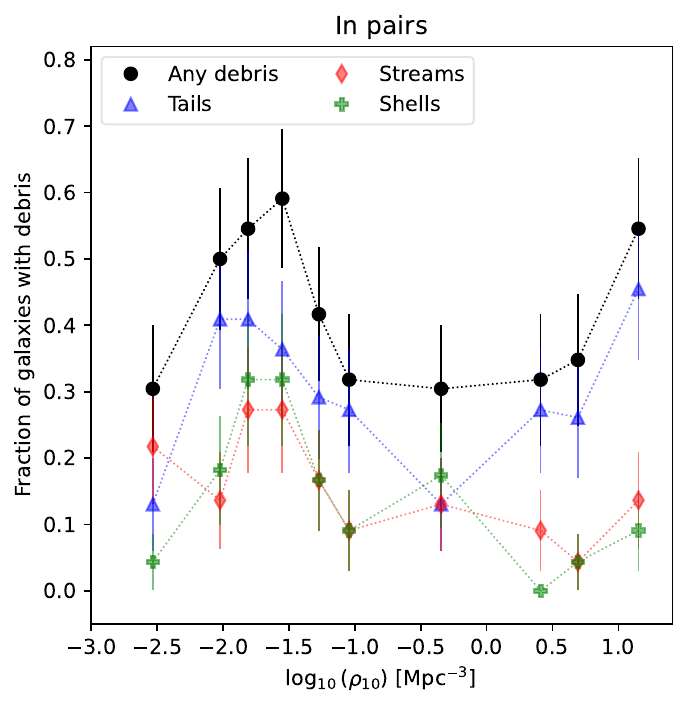}    
   \caption{Fraction of galaxies that have tidal features as a function of the environmental density $\rho_{10}$ (in Mpc$^{-3}$), per bin of $\rho_{10}$. Each bin contains approximately the same number of galaxies (i.e. about 39 galaxies per bin for the top panel, 25 for the bottom left panel and 22 for the bottom right panel). The fraction of galaxies hosting any type of debris is plotted in black,  tails in blue, streams in red and shells in green. The error bars represent the standard deviation in each bin. Galaxies in the Virgo cluster  have $\log_{10}(\rho_{10}) > -0.4$. \textit{Top}: Fraction of galaxies with debris considering all galaxies together. \textit{Bottom left}: Fraction of galaxies with debris only considering isolated (i.e. post-merger) galaxies). \textit{Bottom right}: Fraction of galaxies with debris only considering  galaxies in pairs (i.e. that could be undergoing tidal interactions with a massive companion).}
  \label{fig:results-fraction_galdebris_vs_env}
\end{figure*}

We studied the evolution of the tidal features fraction and luminosity as a function of the large-scale environment parametrised by $\rho_{10}$. The 2-sample KS-test and Mood's test revealed similar distributions and median values of f$_{\rm L,tidal}$ in the field/group and the Virgo cluster. The potential difference between the mean f$_{\rm L,tidal}$ (1.4\% in the field/group and 0.65\% in the cluster) could not be statistically tested because the conditions of application of the test were not verified. In addition, no statistical difference was found for the fraction of galaxies hosting debris, with 35\% in the field/group and 37\% in the Virgo cluster. However, several trends are visible when studying the evolution of this fraction as a function of the environmental density, as presented in the top panel of Figure \ref{fig:results-fraction_galdebris_vs_env}.

The fraction of galaxies hosting debris (Figure \ref{fig:results-fraction_galdebris_vs_env}) exhibits a first well-defined peak between  $-2<\log_{10}(\rho_{10})<-1.5$; it rises from 25\% to a maximum of 50\%.  It then decreases to 30\% before reaching another maximum of 50\% in the highest-density bin. 
 This trend is driven mainly by tidal tails (rising from 10\% to 40\%). 
The first peak is present but less pronounced for galaxies with shells or streams, and their fraction then diminishes  at higher galaxy densities\footnote{We have tested the influence of the binning, and at a $1\sigma$ level, the value of the incidence of debris around $\log_{10}(\rho_{10})= -2$ is always higher (not within the error bars) than the incidence in the lowest-density environment bin. The plateau at intermediate densities is also always visible.}.

The decrease of debris incidence in the lowest environment density bin (about 25\%) was expected, as there are fewer close-by galaxies to interact with. 
The small-scale environment -- on-going collisions generating tidal tails --  explains the two peaks observed in galaxy densities typical of groups and clusters. As shown in the bottom panels of Figure  \ref{fig:results-fraction_galdebris_vs_env}, isolated galaxies do not show any prominent peak for the incidence of tidal features. Similar conclusions hold when considering separately ETGs from LTGs (Figures \ref{fig:appendix-ETGs_fraction_galdebris_vs_env} and \ref{fig:appendix-LTGs_fraction_galdebris_vs_env}). The plateau in the debris incidence at intermediate densities can be related to the influence of the positions of the galaxies inside the cluster, which will be discussed in Section \ref{section: discussion-mass-env-combined}.

This result motivated us to re-analyse the evolution of galaxies with debris as a function of galaxy mass taking into account the small-scale environment. These fractions are presented on the bottom panel of Figure \ref{fig:results-fraction_galdebris_vs_mass}. The trend we report in Section \ref{section:results-effect_galaxy_mass} --  an  increase in the fraction of galaxies with debris with galaxy mass  -- remains valid both for isolated and galaxies in pairs. For isolated galaxies, this fraction ranges between 19\% and 55\% (overall of 31\%), while for galaxies in pairs, it ranges between 27\% and 67\% (overall of 42\%). The overall fractions are statistically different (from the Z-test, with a p-value of 0.016). This increase is mostly driven by tidal tails, with a higher fraction of galaxies with tails among the ones in pairs. No statistically significant difference between isolated and galaxies in pairs is found for streams and shells. The global increase of galaxies hosting debris as a function of galaxy mass is also seen when separating ETGs from LTGs (Figure \ref{fig:appendix-ETGs_LTGs_fraction_galdebris_vs_mass}), with steeper slopes above a mass threshold around $1-3\times 10^{10} M_\odot$.

This analysis confirms, for our SB limit, that the fraction of real post-merger galaxies with debris increases significantly with galaxy mass. Therefore in the recent history (up to a few Gyr), the most massive galaxies in our sample have undergone more merger events. However, it must be noted that tidal tails can also originate from non-merging flybys \citep[e.g.,][]{Gnedin_2003,Sinha_et_al_2012,Mosenkov_et_al_2020}, especially in the Virgo cluster as it will be discussed in Section \ref{section: discussion-mass-env-combined}.

\section{The combined impact of mass and environment}\label{section:results-joint_impact_mass_env}
\begin{figure}
   \centering
   \includegraphics[width=\linewidth]{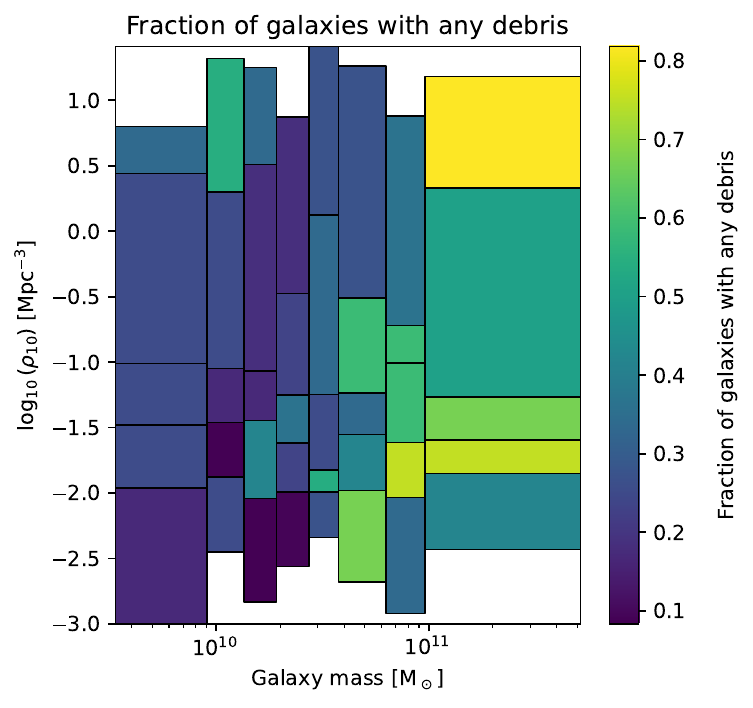}
      \caption{2D-histogram of the fraction of galaxies hosting tidal features as a function of the galaxy mass (in M$_\odot$) and environment density $\rho_{10}$ (in Mpc$^{-3}$). Each 2D-bin contains approximately the same number of galaxies (i.e. 12 galaxies).}
      \label{fig:discussion-hist2D_mass_env}
\end{figure}

In the previous sections, we studied the effect of galaxy mass and environment on the incidence and importance of the LSB stellar structures around galaxies separately. However, it may be argued that mass and environment are correlated, with more massive galaxies residing in denser environments. We illustrate this possible dependence in Figure \ref{fig:discussion-hist2D_mass_env}, which presents 2D-histograms showing the fraction of galaxies with debris as a function of the mass and environment simultaneously. The 2D-bins were computed to contain approximately the same number of galaxies.

There is no evolution with $\rho_{10}$ except in the highest mass bin for the fraction of galaxies with debris at fixed mass. At fixed $\rho_{10}$, there is a trend with galaxy mass for all environment density bins. Hence, this fraction increases more with galaxy mass than with environment density. It must be kept in mind that we only have one example of cluster environment, Virgo, which is still dynamically young.
Since the strongest correlations are found with the mass, our approach of studying the effect of mass and environment separately seems justified.

\section{Trends as a function of the internal kinematics of the host galaxy}\label{section:results-effect_internal_kinematics}
In previous sections, galaxy morphologies were used to separate our sample between ETGs and LTGs. However, other classification schemes exist, such as internal stellar kinematics, which may be a more robust and fundamental parameter to distinguish the various types of galaxies.
Most ETGs show regular rotation patterns (fast rotators, FRs) while others do not show any sign of rotation or complex kinematic features (slow rotators, SRs) \citep[e.g.,][]{Emsellem_et_al_2007,Emsellem_et_al_2011}. Separating SRs from FRs is essential when studying the evolution of ETGs, as a number of numerical simulations predict that ETGs should initially form as FRs and later transform into SRs through mergers \citep[e.g.,][]{Naab_et_al_2014,Penoyre_et_al_2017,Lagos_et_al_2018,Lagos_et_al_2022}. 
\cite{Bilek_et_al_2022b} combined kinematic information from the ATLAS$^{3D}$ sample of ETGs and  merger-sensitive features (such as tidal features, kinematically distinct cores, stellar ages)  to study the role of mergers in the transformation of FRs into SRs for the MATLAS sample \citep{Duc_et_al_2015,Bilek_et_al_2020}. According to their work, internal kinematics was implemented early (around $z=2$) through multiple minor wet mergers. 

Here, we extend the work of \cite{Bilek_et_al_2022b} \citep[and previous works from the ATLAS$^{3D}$ collaboration, e.g.,][]{Duc_et_al_2011}, and investigate the relationship between kinematics and the late assembly history of ETGs. We adopt a quantitative approach to the topic, and further probe denser environments by including ETGs in the Virgo cluster. 

\begin{figure}
    \centering
    \includegraphics[width=0.4\textwidth]{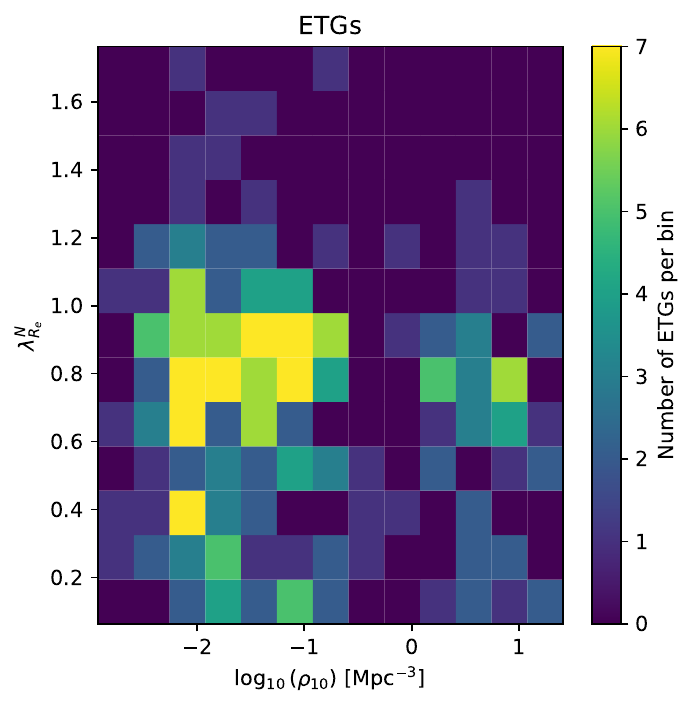}
    \caption{2D histogram of the distribution of the number of ETGs as a function of the rotational support, $\lambda_{\rm R_e}^{N}$, and of the environment density, $\rho_{10}$, in Mpc$^{-3}$. Galaxies in the Virgo cluster have $\log_{10}(\rho_{10}) > -0.4$. Galaxies with $\lambda_{\rm R_e}^{N}$ < 0.31 (resp. > 0.31) are considered SRs (resp. FR).}
    \label{fig:results-hist2d_ETG_lambda_rho10}
\end{figure}

Following \cite{Emsellem_et_al_2011} and \cite{Bilek_et_al_2022b}, we based our analysis on the estimated rotational support, $\lambda_{\rm R_e}^{N} = \lambda / \sqrt{\epsilon}$, where $\lambda$ quantifies the degree of ordered rotation of the galaxy and $\epsilon$ the apparent ellipticity. This also includes 2-$\sigma$ galaxies, which are characterised by two off-centre, but symmetric, peaks in the velocity dispersion, which lie on the major-axis of the galaxy \citep{Krajnovic_et_al_2011}. $\lambda_{\rm R_e}^{N}$ is measured within one effective radius $R_e$. Galaxies with $\lambda_{\rm R_e}^{N}$ < 0.31 (resp. > 0.31) are considered SRs (resp. FR). With such a definition, 15\% of our ETGs are SRs. From the 2-sample KS-test and Mood's test, SRs in our sample are statistically significantly more massive than FRs. 
The distribution of our ETGs in the plane $\lambda_{\rm R_e}^{N}$ versus environment density is shown in Figure \ref{fig:results-hist2d_ETG_lambda_rho10}. ETGs in our sample are preferentially located in the group environment (median $\log_{10}(\rho_{10}) = -1.4$) with a relatively high rotational support (median $\lambda_{\rm R_e}^{N} = 0.76$).

\begin{figure*}
    \centering
    \includegraphics[width=\linewidth]{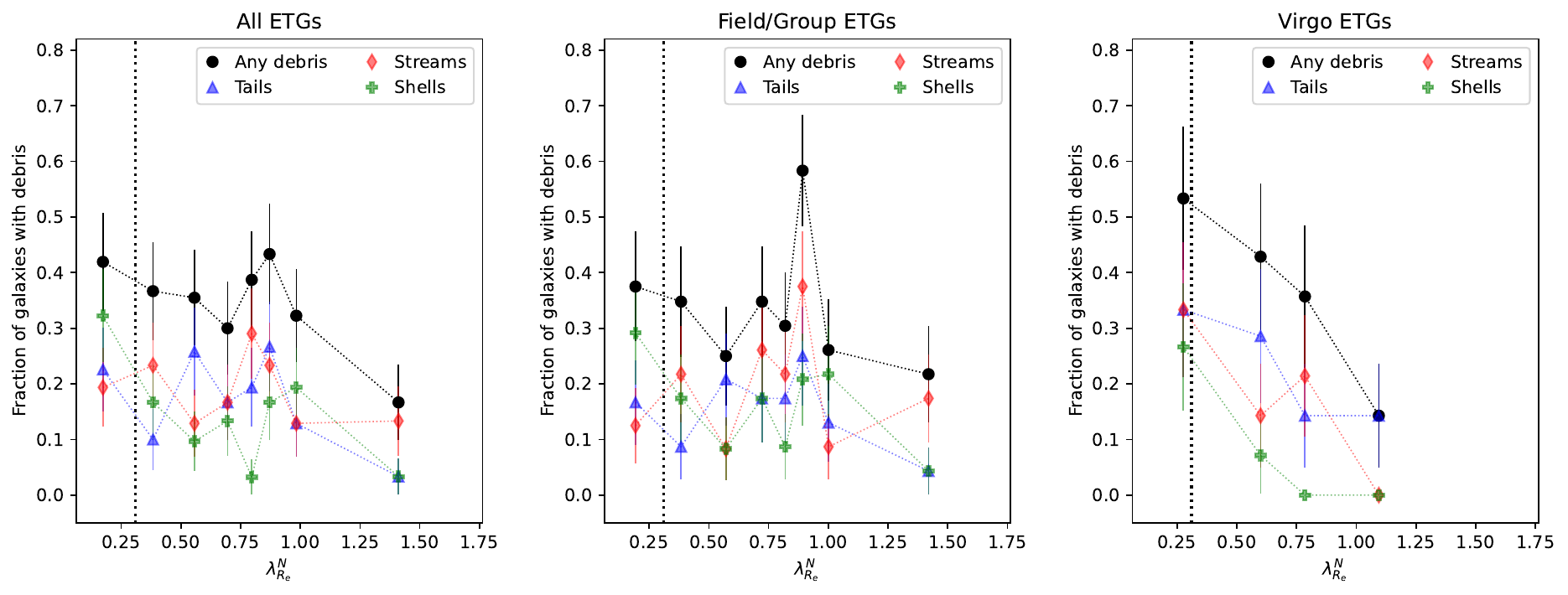}
    \caption{Scatter plot of the fraction of galaxies that have tidal features as a function of the rotational support, $\lambda_{\rm R_e}^{N}$. Each bin contains approximately the same number of galaxies (i.e. 30 galaxies per bin for the left plot, 23 for the middle plot and 14 for the right plot). The fraction of galaxies hosting any type of debris is plotted in black, tails in blue, streams in red and shells in green. The error bars represent the standard deviation in each bin. The vertical dotted line separates SRs from FRs. Such fractions are presented for all ETGs (\textit{left}), field/group ETGs (\textit{middle}) and Virgo ETGs (\textit{right}). }
    \label{fig:results-frac_galdebris_lambda}
\end{figure*}

We display the fraction of ETGs hosting debris as a function of the rotational support, $\lambda_{\rm R_e}^{N}$, in Figure \ref{fig:results-frac_galdebris_lambda}.
Globally, when considering all SRs versus all FRs, 42\% of SRs host tidal debris against 33\% for FRs, a difference which is not statistically significant according to the Z-test. However if we separate galaxies in Virgo versus in the field/group, a decrease in the fraction of galaxies with tidal features with $\lambda_{\rm R_e}^{N}$ is especially clear for Virgo ETGs. In addition, galaxies around $0.7<\lambda_{\rm R_e}^{N}<1$ tend to have a higher fraction of tidal debris, a trend best visible in low galaxy density environments. 
As ETGs with this range of $\lambda_{\rm R_e}^{N}$ are preferentially located in the group environment (as seen in Figure \ref{fig:results-hist2d_ETG_lambda_rho10}), this increase could be triggered by multiple interactions within the group, similar to the one observed in Figure \ref{fig:results-fraction_galdebris_vs_env}.

\begin{figure*}
    \centering
    \includegraphics[width=0.75\linewidth]{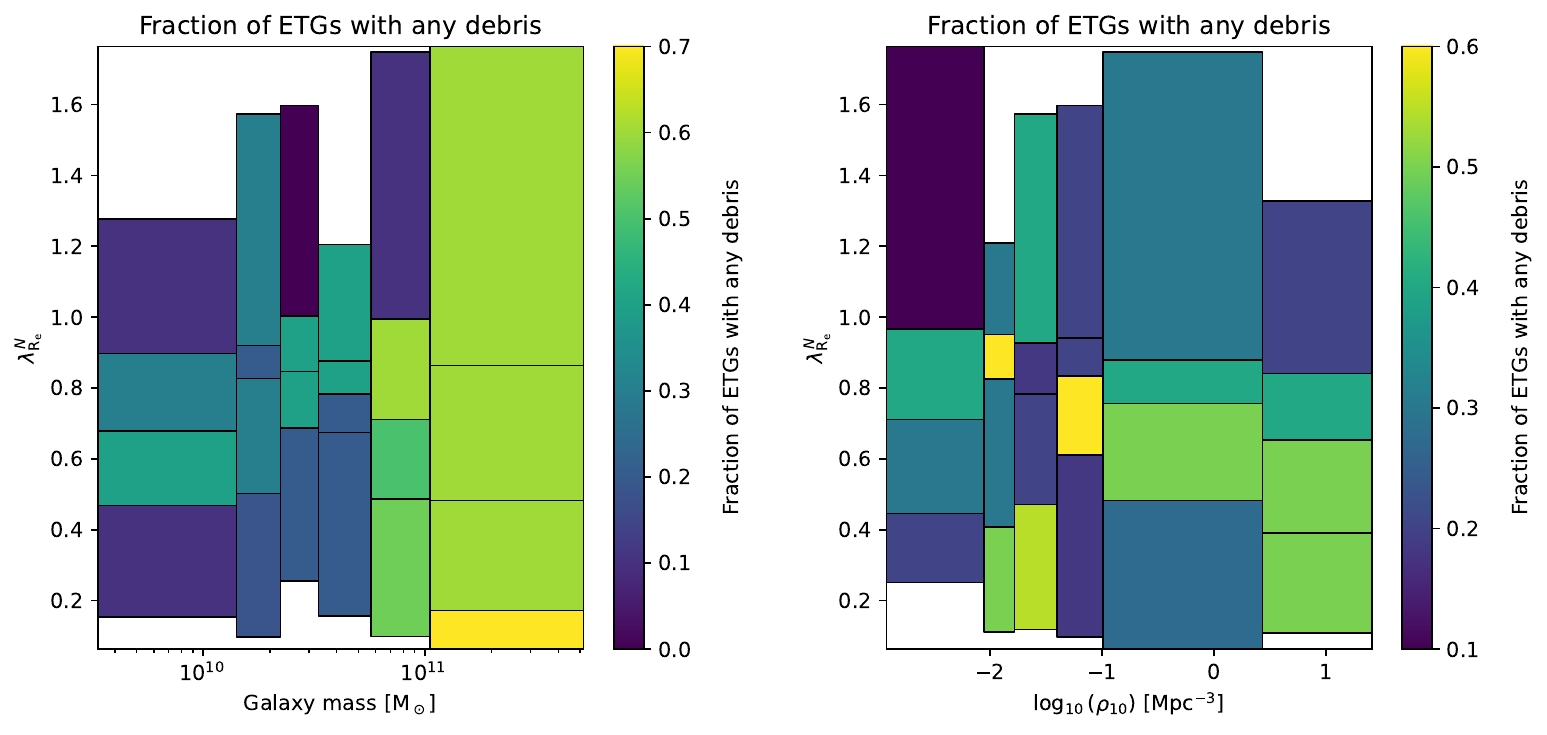}
    \caption{2D-histograms of the fraction of ETGs hosting tidal features as a function of the degree of ordered rotation, $\lambda_{\rm R_e}^{N}$, and the galaxy mass (\textit{left}, in M$_\odot$) and the environment density, $\rho_{10}$, (\textit{right}, in Mpc$^{-3}$).
    Each 2D-bin contains approximately the same number of galaxies (i.e. about 10 galaxies). Galaxies with $\lambda_{\rm R_e}^{N}$ < 0.31 (resp. > 0.31) are considered slow rotators (resp. fast rotators).}
    \label{fig:results-2Dhist_kinematics}
\end{figure*}

In the top panel of Figure \ref{fig:results-2Dhist_kinematics}, we investigated whether we could separate the effect of mass and environment from internal kinematics through 2D histograms. Indeed, slow rotators are statistically more massive than fast rotators \citep[e.g.,][]{Khochfar_et_al_2011}.  For the fraction of ETGs hosting debris, there is a correlation with mass, but the trend with $\lambda_{\rm R_e}^{N}$ is less clear, in part because of our low number of SRs. The trends between environment density and $\lambda_{\rm R_e}^{N}$ seem less clear, with higher fractions of ETGs with debris for lower values of ordered rotation.  
A proper disentangling between the effect of the mass, environment and internal kinematics would require  a study of the correlations at fixed mass and environment density, such as the one carried out by \cite{Bilek_et_al_2022b}. They found that the trends of tidal features stopped being statistically significant after accounting for the effect of mass. They concluded that the kinematic state was established at a high redshift, but the SRs keep experiencing more mergers until today.

\section{Discussion}\label{section:discussion}
\subsection{Incidence and luminosity of LSB features}
All our results are reported for a SB limit  $\mu_{r,lim (3\sigma, 10 \arcsec\times 10 \arcsec)}$ of $\thicksim 29$ mag$\,$arcsec$^{-2}$.
In Section \ref{section:results-census_features} we found that 36\% of our galaxies host tidal features (23\% host tails, 15\% host streams and 12\% host shells). A direct comparison to the literature is not straightforward. The properties of the galaxies (e.g., distances, prominence of the tidal features) and properties of the studies (e.g., definitions of the types of tidal features, surface brightness limits, detection methods, image processing techniques, sample selection) differ from one study to the other \citep[e.g.,][]{Atkinson_Abraham_Ferguson_2013,Hood_et_al_2018}.

This results in large discrepancies in the reported fraction of tidal features in nearby galaxies ranging from a few percent up to 70\% \citep[e.g.,][]{Malin_and_Carter_1983,Schweizer_and_Seitzer_1988,van_Dokkum_2005,Tal_et_al_2009,Bridge_et_al_2010,Nair_and_Abraham_2010,Kaviraj_2010,Miskolczi_et_al_2011,Kim_et_al_2012,Sheen_et_al_2012,Adams_et_al_2012,Atkinson_Abraham_Ferguson_2013,Duc_et_al_2015,Hood_et_al_2018,Bilek_et_al_2020,Jackson_et_al_2021,Vazquez-Mata_et_al_2022,Yoon_et_al_2023,Rutherford_et_al_2024,Skryabina_et_al_2024,Miro_Carretero_et_al_2024a}. 
On the simulation side, \cite{Pop_et_al_2018} produced a census of shells in the Illustris simulation and found an incidence of 20-30\% for shells, which is higher than our finding, but they did not apply any SB limit cut. Similarly, \cite{Valenzuela_et_al_2024} found that 10–30\% of their high-mass simulated galaxies host tidal features, with a higher fraction for streams (29\%). This higher value can be explained by their use of 3D information to identify tidal features and the fact that their galaxies are more massive than those in our sample. \cite{Martin_et_al_2022} also produced a census of tidal features in simulated deep images considering different depths. They found fractions roughly similar to ours (between a few per cent and less than 20\%) for an SB limit of 29 mag$\,$arcsec$^{-2}$. \cite{Vera-Casanova_et_al_2022} used mock images from Auriga simulations at varying depths, and found that 30–40\% of their Milky Way-like galaxies host streams at a surface brightness limit of 29 mag$\,$arcsec$^{-2}$. This fraction is higher than ours, likely due to several factors: the simulations lack background noise and artefacts, the galaxy sample is smaller, and a detection is recorded as soon as the first signs of streams appear, whereas our study requires higher confidence levels to delineate streams. The variability in the stream fraction between different simulations has also been reported by \cite{Miro-Carretero_et_al_2024b}, who found fractions ranging between 10-35\% for a SB limit of 29 mag$\,$arcsec$^{-2}$ in three simulations.

Tails and streams  account for 2-4\% of the total galaxy luminosity (Section \ref{section:results-luminosities}). Even though the definitions of tidal features differ (as mentioned above), similar numbers are found in the literature. \cite{Jackson_et_al_2021} estimated the stellar material in the outskirts of their massive central galaxies, mostly in the form of tidal and merger features, represent a few per cent of the total stellar mass (8\% outside 2R$_e$, which likely includes a part of what we defined as halo). \cite{Huang_and_Fan_2022} found their f$_{\rm L,tidal}$ lower than 1\% for tidal features around their massive ETGs in residual HSC-SSP images.
In simulations, \cite{Martin_et_al_2022} found that the flux in tidal features ranges from 1 to 10\% for the most massive galaxies. Their values are higher as they retrieve all the particles associated to a feature since the ground truth and the fraction of stars from accreted origin is known.

As observed in \cite{Sola_et_al_2022}, we find that streams are fainter than tails by 0.7 mag, a statistically significant difference confirmed by Mood's test and a 2-sample KS-test. This may be related to different survival lifetimes \citep[e.g.,][]{Mihos_1995,Ji_et_al_2014,Mancillas_et_al_2019}, with streams tending to remain visible longer than tails.
There is no statistically significant difference between the SB of tails associated to on-going interactions and those of isolated galaxies, as the p-values of the statistical test are larger than the 5\% threshold.

No features are observed below 27.8 mag$\,$arcsec$^{-2}$ which is about 1 mag brighter than the SB limit. This is related to the difference in the measure of the SB limit (in $10\times10 \arcsec$ boxes) and the SB of the extended, diffuse tidal features that display brightness fluctuations, as well as the fact that the annotation has been done visually and the eye is not able to detect the faintest structures. In addition, we report the median values, but some parts of the tidal features are fainter than others. This difference must be considered when comparing observations to simulations.

\subsection{The combined effect of  mass and environment} \label{section: discussion-mass-env-combined}
As presented in Sections  \ref{section:results-effect_galaxy_mass}, \ref{section:results-effect_environment} and \ref{section:results-joint_impact_mass_env}, galaxy mass is the dominant effect that impacts the frequency of tidal features, a result also highlighted in previous publications. 
For instance, \cite{Atkinson_Abraham_Ferguson_2013} noted a mass-dependency, with an increased fraction of galaxies having tidal features (linear features, shells and fans of stellar light) for galaxy masses $>10^{10.5}=3.16\times10^{10}M_\odot$, which is very close to the mass threshold (i.e., a sharp increase of some properties above a given stellar mass) we observe. Likewise, \cite{Duc_et_al_2015} and \cite{Bilek_et_al_2020} reported an increased fraction of ETGs with shells, streams and disturbed isophotes with increasing galaxy mass, mostly for galaxies with a mass $>10^{11}M_\odot$. Other studies, such as \cite{Yoon_and_Lim_2020}, find a strong correlation between the fraction of ETGs with features and increasing galaxy mass: 30-40\% of their massive ($M_{\rm dyn}>10^{11.4}M_\odot$) galaxies have tidal features, while 2-5\% of their less massive ($M_{\rm dyn}<10^{10.4}M_\odot$) ETGs show features. Similarly, \cite{Vazquez-Mata_et_al_2022} found the fraction of galaxies with any type of debris to increase with stellar mass, from 10\% to 30\% for their most massive galaxies. The correlation between stellar mass and the presence of tidal features is also noted by  \cite{Skryabina_et_al_2024}.

On the simulation side, \cite{Martin_et_al_2022} also noted a slight increase in the fraction of galaxies with shells, streams or tails with galaxy mass for an SB limit of 29 mag$\,$arcsec$^{-2}$. In four different numerical simulations, \cite{Khalid_et_al_2024} found a similar increase of galaxies hosting tails and streams with stellar mass, ranging from a few per cent to about 20\% for stellar masses around $10^{10} M_\odot$.
 \cite{Huang_and_Fan_2022} noted an increase in the fraction of flux in tidal features,  f$_{\rm L,tidal}$,  from 0.5\% for stellar masses around $10^{11} M_\odot$ to 1\% for masses higher than $10^{12} M_\odot$. \cite{Martin_et_al_2022} reported a substantial increase in f$_{\rm L,tidal}$ with galaxy mass, above a stellar mass threshold of $10^{10.1} M_\odot$.

The presence of such a mass threshold, above which the increase of the aforementioned quantities are steeper ($3\times10^{10}M_\odot$ for the fraction of galaxies with debris, $4-7\times10^{10} M_\odot$ for f$_{\rm L,tidal}$) has also been reported in the literature: between a few $10^{10}$ and $10^{11}M_\odot$ \citep[e.g.,][]{Atkinson_Abraham_Ferguson_2013,Duc_et_al_2015,Bilek_et_al_2020,Yoon_and_Lim_2020,Martin_et_al_2022,Vazquez-Mata_et_al_2022,Huang_and_Fan_2022}. A threshold of $10^{10.81} M_\odot$  has also been noted by \cite{Lim_et_al_2023} through the census of the  globular cluster (GC) distributions.  They interpret this as high-mass galaxies having accreted their GC systems while lower-mass galaxies have formed them in situ.

This approximate mass threshold also coincides with the galaxy mass where merger events become the dominant factor of galactic growth according to numerical simulations \citep[e.g.,][]{Stewart_et_al_2008,Rodriguez-Gomez_et_al_2015}. Likewise, \cite{Kauffmann_et_al_2003} noted a change in galaxy properties around $3\times10^{10} M_\odot$, which matches our findings, with lower-mass galaxies having young stellar populations, disc-like structures, and low surface mass densities.
These findings are consistent with the idea that the mass growth of the most massive galaxies is dominated by mergers \citep[e.g.,][]{Newman_et_al_2012,Hilz_et_al_2013,Robotham_et_al_2014,Rodriguez-Gomez_et_al_2016,Vulcani_et_al_2016,Hill_et_al_2017a,Hill_et_al_2017b,Nevin_et_al_2023,Guzman-Ortega_et_al_2023}, as gas accretion may no longer be possible (due to too hot haloes that prevent the fresh infall of gas), hence an increased fraction and luminosity of tidal features. \cite{Keres_et_al_2005} and \cite{Dekel_and_Birnboim_2006} noted that around a similar mass threshold, the golden mass, lower-mass galaxies are dominated by cold accretion (i.e., gas accreted from cold, dense intergalactic filaments) and higher-mass galaxies are dominated by hot accretion (i.e., gas shock-heated to high temperature that later cools). Unlike massive ETGs, LTGs tend to have a smoother accretion history, mostly through gas accretion \citep[e.g.,][]{Sancisi_et_al_2008, Bilek_et_al_2022b}. Indeed, if the LTGs had undergone major mergers, they would eventually have turned into ETGs \citep[although gas-rich mergers can sometimes produce a spiral remnant, e.g.,][]{Springel_and_Hernquist_2005}; hence, the LTGs visible today should have had a relatively quiet evolution in the past few gigayears. 
This seems to be supported by Figure \ref{fig:results-fraction_galdebris_ETGs_LTGs_vs_mass} in which LTGs show a lower occurrence of shells and streams (resulting from mergers) but more tidal tails (more common for rotation-dominated systems and that can also originate from non-merging flybys). 

Furthermore, all our results are obtained for an approximate $r$-band surface brightness limit $\mu_{r,lim (3\sigma, 10 \arcsec\times 10 \arcsec)}$ of 29 mag$\,$arcsec$^{-2}$. One may wonder whether the increase in the fraction of galaxies with debris with increasing mass (Figure \ref{fig:results-fraction_galdebris_vs_mass}) could be explained by the fact that less massive galaxies would have fainter hence non-detected tidal features, rather than by an effect of galaxy mass. Numerical simulations can help disentangle between both effects. \cite{Martin_et_al_2022} investigated the detectability and fraction of flux in tidal features as a function of surface brightness limit and stellar mass. From their Figure 12, more massive galaxies have tidal features which are more easily detected. For an SB limit of 29 mag$\,$arcsec$^{-2}$, the less massive galaxies ($10^{9.5}-10^{10} M_\odot$) have no detectable features, while for galaxies above $10^{10.5} M_\odot$ more than 20\% of the area of features is detectable. Therefore, in our study the actual fraction of galaxies with debris should be higher for all mass ranges, not only for the less massive ones. In addition, Figure 8 of \cite{Martin_et_al_2022} shows an increase in the fraction of flux in tidal features with stellar mass for all depths, although the normalisation and scatter of the relations strongly depend on depth. For an SB limit of 29 mag$\,$arcsec$^{-2}$, the flux in features increases from 0.04\% for $5\times 10^{9} M_\odot$ galaxies up to 7\% for $3.2\times 10^{11} M_\odot$ ones. \cite{Martin_et_al_2022} conclude that  more massive galaxies have tidal features which are both more frequent and brighter (in flux). 

This numerical prediction (more flux in tidal features around more massive galaxies) aligns with our finding in Section \ref{section:results-effect_galaxy_mass}. However, there might still be a detection bias: if the surface brightness (and not only total luminosity) of tidal features actually scales with galaxy mass, then due to our limited depth we would miss the faintest features around the lower-mass galaxies. We examined how the median surface brightness of detected tidal features varies with host stellar mass and we report no significant trend. Since tidal feature detectability is driven primarily by surface brightness limits of the survey, the lack of any correlation between mass and surface brightness indicates that completeness does not worsen for lower-mass galaxies. We conclude that the increasing fraction of galaxies hosting features with galaxy mass is genuinely driven by mass, rather than by a detection bias.

In this paper, we report no strong influence of the large-scale environment (distinguishing the Virgo cluster from the field/group) on the incidence of tidal features (see Section \ref{section:results-effect_environment}), except in the group environment probed by our sample. 
A possible increase in the fraction of galaxies with tails and shells in the group environment was also seen by \cite{Bilek_et_al_2020}. Previous studies reported decreasing fractions of shells in higher-density environments \citep[e.g.,][]{Malin_and_Carter_1983}, which was explained by the fact that shells originate mostly from radial mergers with low-velocity encounters \citep[e.g.,][]{Quinn_1984}, unlike the high-velocity encounters in clusters. Indeed, in massive massive clusters such as Virgo the velocity dispersion is high \citep[about 1000 km$\,$s$^{-1}$,][]{Gunn_and_Gott_1972}, thus gravitational interactions are short living phenomena \citep{Boselli_and_Gavazzi_2006}. The probability of producing tidal tails or other LSB features should be very low. Here, we do not see any statistically significant trend for shells and streams. However, there is a net increase in the fraction of galaxies with tails in the cluster (Figures \ref{fig:results-fraction_galdebris_vs_env}, \ref{fig:appendix-ETGs_fraction_galdebris_vs_env} and \ref{fig:appendix-LTGs_fraction_galdebris_vs_env}). This suggests that although tidal feature evaporation was expected in clusters, the Virgo cluster environment is not unfavourable to the formation of young tidal tails. As previously mentioned, the Virgo cluster is still dynamically young \citep[e.g.,][]{Binggeli_et_al_1987,Boselli_et_al_2014} and assembling, so it is possible that tidal tail destruction would be compensated by the quick formation of new tails, including a potential enhancement due to non-merging flybys, 
making the distinction between these competing processes complicated. Deriving broader conclusions on the impact of galaxy clusters on tidal features would require studying other clusters, including virialised ones such as Coma, which would be worth conducting in future works.

Another factor to consider is the location of the galaxies inside the cluster itself, as strong dependencies have been noted in the literature between the positions and some galaxy properties \citep[e.g.,][]{Gnedin_2003, Mihos_2004,Boselli_and_Gavazzi_2006,Janowiecki_et_al_2010,Adams_et_al_2012, Boselli_et_al_2014,Perez_montano_et_al_2024}. The study the evolution of the fraction of galaxies with debris as a function of the clustercentric distance revealed the same trend, i.e. that small-scale environment seems responsible for the formation of tidal features, especially tails. The small-scale environment is indeed closely tied to our definition of tidal tails, as their formation typically implies the past or ongoing presence of nearby massive companions.

An extension of this environmental study to the influence of the large-scale structure of the Universe on tidal features could be valuable and will be the topic of another paper, as e.g.,  \cite{Perez_montano_et_al_2024} showed that large-scale structure has a minimal impact on determining the LSB nature of galaxies in their sample.

\subsection{The effect of internal kinematics}
In Section \ref{section:results-effect_internal_kinematics}, we found increased fractions of SRs hosting shells compared to FRs. Other works have investigated the links between tidal features and internal kinematics. For instance, \cite{Rutherford_et_al_2024} used streams and tails around SRs and FRs in the SAMI Galaxy Survey using deep HSC data, in order to trace their dynamical evolution. Likewise, \cite{Yoon_et_al_2024} studied the fraction of galaxies with tidal features as a function of the kinematics of 1244 ETGs.  The correlations between tidal features and rotational supported in simulations was also investigated by \cite{Valenzuela_et_al_2024}. The higher fraction of shells around SRs than FRs was also reported by \cite{Rutherford_et_al_2024}. Similarly, \cite{Yoon_et_al_2024} found that half of their ETGs with shells are SRs, and that ETGs with tidal features have reduced rotational support compared to ETGs without such features. They also found that ETGs with low $\lambda_{\rm R_e}^{N}$ preferentially host shells compared to tails and streams, which is comparable to what we observed except in the group environment where these fractions can be increased due to close interactions. These observational results are consistent with the higher fraction of shells around SRs in simulations, as reported by \cite{Valenzuela_et_al_2024} who showed that most galaxies with shells became SRs at early times, in agreement with \cite{Bilek_et_al_2022b}.

Finally, it must be noted that other effects than mergers act on galaxies in clusters. The numerous, high-velocity flybys \citep[e.g.,][]{Moore_et_al_1996}, particularly at near the cluster center, contribute to stretch the outer galactic material further away. \cite{Boselli_et_al_2014} found for ETGs in the Virgo cluster that the most massive galaxies are SRs, and the lower-mass SRs preferentially located in the higher-density substructures of the cluster. They outline that the massive systems likely result from major mergers at early times and for the less massive SRs that they have been Virgo members since the cluster formation. Therefore, since SRs are preferentially located towards the core and highest-density regions of the Virgo cluster, they would be more prone to undergo the high-velocity flybys than FRs and hence have more extended haloes.

\section{Summary and conclusions}\label{section:conclusion}
Low surface brightness stellar structures around galaxies hold crucial clues about their late assembly history of the host galaxies. 
In this paper, we investigated the relationships between several quantitative properties of LSB tidal features and the host galaxy properties (mass, environment, and internal kinematics) for a sample of 475 nearby massive galaxies.

Using the annotation tool \textit{Jafar} \citep{Sola_et_al_2022}, we have manually delineated LSB features around our 475 galaxies in deep images from four CFHT surveys: MATLAS, CFIS/UNIONS, VESTIGE and NGVS. Our galaxy sample, drawn from the ATLAS$^{3D}$ reference catalogue \citep{Cappellari_et_al_2011}, contains similar numbers of ETGs and LTGs and probes the field, group and Virgo cluster. From our compiled database of 11832 annotated structures, including 199 tails and 100 streams, we quantitatively characterised LSB features' properties and coordinates in Table \ref{table:appendix-list_features} and in Appendix \ref{section:appendix-coordinates_features}. The results we report are obtained for a surface brightness limit of about 29 mag$\,$arcsec$^{-2}$.
Our catalogue of LSB features and their properties can be used as a baseline for future studies aiming to characterise tidal features and compare them to numerical simulation predictions. Our annotated structure dataset also provides labelled data for training machine learning algorithms to automatically find tidal features in deep images \citep{Richards_et_al_2024}.

We summarise our results below.
\vspace{11pt}

\textbf{ a) Incidence and luminosity of LSB features }
\begin{itemize}
    \item 36\% of our galaxies display one or more tidal features: 23\% host tidal tails, 15\% host streams and 12\% host shells (Section \ref{section:results-census_features}).
    \item Tidal tails and streams account for 2-4\% of the total galaxy luminosity  (Section \ref{section:results-census_features}).  
    \item Streams are fainter than tails, with a median surface brightness of 26.8 and 26.1 mag$\,$arcsec$^{-2}$, respectively. The faintest tidal feature is about one magnitude brighter than the SB limit.
\end{itemize}

The impact of mass, environment and internal kinematics on the incidence and luminosity of the tidal features were studied separately, though they might be linked. This issue is discussed in Section \ref{section:results-joint_impact_mass_env}.

\vspace{11pt}

\textbf{b) Trends with galaxy mass }
\begin{itemize}
    \item The fraction of galaxies hosting debris increases with galaxy mass, doubling from around 25\% to 60\% for the highest-mass galaxies. This trend is visible for each type of debris (Figure \ref{fig:results-fraction_galdebris_vs_mass}). 
    \item More massive galaxies host more luminous tidal features (Section \ref{section:results-effect_galaxy_mass}).
    \item There is a mass threshold around $4-7\times10^{10}M_\odot$ above which the slopes of the trends mentioned above are steeper (Figure \ref{fig:results-fraction_galdebris_vs_mass}).
\end{itemize}

\vspace{11pt}

\textbf{c) Trends with the environment }

\begin{itemize}
    \item The large-scale environment (field/group versus Virgo cluster)  does not impact the fraction of galaxies hosting tidal features (Figure \ref{fig:results-fraction_galdebris_vs_env}).
    \item The presence of a nearby companion (in projected separation) leads to increased fractions of galaxies with tidal tails in the field/group and in the Virgo cluster (Figure \ref{fig:results-fraction_galdebris_vs_env}). 
\end{itemize}

\textbf{ d) Trends with internal kinematics  }
\begin{itemize}
    \item The fraction of slow rotators (42\%) hosting debris is higher than that of fast rotators (33\%).   
\end{itemize}
These trends could, however, be due to the mass effect previously identified, as slow rotators are more massive than fast rotators. A proper analysis is needed to study the correlations at fixed mass and environment density.

To summarise, our quantitative analyses of LSB features around hundreds of massive galaxies enabled us to obtain hints about their recent assembly history. Our findings are consistent with the hierarchical build-up of galaxies, with massive galaxies having assembled the majority of their mass through mergers rather than by gas accretion. Mass, rather than the environment, is the dominant factor that affects tidal features.  More massive galaxies have undergone more mergers and thus show more signs of tidal disruption, and this is especially true for ETGs.
The presence of a mass threshold, also noted in the literature, is consistent with a picture where galaxy growth is dominated by gas accretion for low-mass galaxies and by mergers for high-mass galaxies, where gas accretion may no longer be possible.

Finally, although tidal features are the most prominent signs of interactions, their lifetime is limited, and they will eventually phase-mix and contribute to the diffuse light of the extended stellar halo. A detailed study of the surface brightness profiles, luminosity and colours of stellar haloes, as well as their relations to tidal features, will be the focus of a future paper. This will enable us to probe galaxies' assembly history over longer periods than with tidal features alone.
Another future avenue is the comparison of observations to the predictions of numerical simulations, and our findings can be used for that purpose. Such work will require larger samples of LSB features, for which efficient automated methods still have to be developed.

\section*{Acknowledgements}
We thank the anonymous referee for a thorough review and comments that helped shaping this article.
E.S. would like to thank Eric Emsellem for interesting discussions and for his thorough review of the paper which helped improve it. We thank Ariane Lan\c{c}on for her comments on this work.  
We thank Maarten Baes, Crescenzo Tortora and Ivana Ebrova for providing mock images from the Illustris TNG50 and TNG100 simulations.

E.S is grateful to the Leverhulme Trust for funding under the grant number RPG-2021-205. 
M.P. is supported by the Academy of Finland grant n:o 347089. O.M. is grateful to the Swiss National Science Foundation for financial support under the grant number PZ00P2\_202104. M.B. is grateful for the visiting professorship from the University of Vienna. R.H. acknowledges funding from the Italian INAF Large Grant 12-2022.
M.B. was supported by the Ministry of Science, Technological Development and Innovation of the Republic of Serbia under contract no. 51-03-136/2025-03/200002 with the Astronomical Observatory of Belgrade. F.D. acknowledges support from CNES.

This research has made use of the SIMBAD database, operated at CDS, Strasbourg, France, and of "Aladin sky atlas" developed at CDS, Strasbourg Observatory, France.
This work is based on data obtained as part of the Canada-France Imaging Survey, a CFHT large program of the National Research Council of Canada and the French Centre National de la Recherche Scientifique. Based on observations obtained with MegaPrime/MegaCam, a joint project of CFHT and CEA Saclay, at the Canada-France-Hawaii Telescope (CFHT) which is operated by the National Research Council (NRC) of Canada, the Institut National des Science de l’Univers (INSU) of the Centre National de la Recherche Scientifique (CNRS) of France, and the University of Hawaii. This research used the facilities of the Canadian Astronomy Data Centre operated by the National Research Council of Canada with the support of the Canadian Space Agency. This research is based in part on data collected at Subaru Telescope, which is operated by the National Astronomical Observatory of Japan. We are honored and grateful for the opportunity of observing the Universe from Maunakea, which has the cultural, historical and natural significance in Hawaii. Pan-STARRS is a project of the Institute for Astronomy of the University of Hawaii, and is supported by the NASA SSO Near Earth Observation Program under grants 80NSSC18K0971, NNX14AM74G, NNX12AR65G, NNX13AQ47G, NNX08AR22G, YORPD20\_2-0014 and by the State of Hawaii.  
This research was supported by the International Space Science Institute (ISSI) in Bern, through ISSI International Team project \#534.

\section*{Data Availability}

To ensure the reproducibility of this work, we share in Appendix \ref{section:appendix-coordinates_features} the properties of the individual 199 tidal tails and 100 streams used in this study. We complement this by providing a file in the DS9 Region format that contains the coordinates of the contours of all our LSB features, including tails, streams, shells, haloes and inner galaxies. These can be used to create the mask of each annotation. The MATLAS images are publicly available and can be accessed e.g. through Aladin.
The CFIS/UNIONS, NGVS, and VESTIGE surveys are all publicly available via CADC in the standard format optimized for compact source science \citep{elixir}, without low surface brightness processing — such processing has been undertaken within dedicated collaborations. While UNIONS plans to make its data, including the LSB processing, publicly available in the near future, access to the LSB products for NGVS and VESTIGE currently requires direct contact with the respective PIs.


\bibliographystyle{mnras}
\bibliography{bibliography} 



\appendix

\onecolumn

\section{Note on the surface brightness limit}\label{section:appendix-sblim}
We indicated in Section \ref{section:data} the surface brightness limits of MATLAS, CFIS and VESTIGE as reported in the literature. However these values were not computed in a consistent manner, which can lead to significance differences that are due to the method and not the image. Here, we recompute the surface brightness limits of the $r-$band images for these three surveys using the method described in Appendix A of \cite{Roman_et_al_2020}. 

As the value of the SB limit strongly depends on the image and the presence of contamination sources (e.g., cirrus, bright stars, ghost reflections), we visually select three `clean' images per survey, at the original pixel scale. We mask all sources using SExtractor. We then randomly select fifty cutouts of varying size per masked image. For each cutout, we compute the histogram of the pixel values and fit a Gaussian to it. The resulting standard deviation $\sigma$ is used to estimate $\mu_{lim (3\sigma, 10 \arcsec\times 10 \arcsec)}$, the surface brightness limit at the $3\sigma$ level on an angular scale of $10 \arcsec \times 10 \arcsec$, as: 
\begin{equation}
    \mu_{lim (3\sigma; 10 \arcsec\times 10 \arcsec)} = -2.5 \times \log_{10} \left( \frac{3\sigma}{pix\times 10} \right) + zp
\end{equation}
where $pix$ is the original pixel size and $zp$ is the zeropoint.

We find that the re-computed $\mu_{lim (3\sigma, 10 \arcsec\times 10 \arcsec)}$ for MATLAS is around 29.2 mag$\,$arcsec$^{-2}$; it is of 28.7 mag$\,$arcsec$^{-2}$ for CFIS and 29.1 mag$\,$arcsec$^{-2}$ for VESTIGE. For MATLAS, this value is slightly higher than the quoted 28.9 mag$\,$arcsec$^{-2}$  (Cuillandre, private communication), and likewise for CFIS (with a quoted 28.3 mag$\,$arcsec$^{-2}$, Cuillandre, private communication). A larger difference is seen for VESTIGE with a quoted 27.2 mag$\,$arcsec$^{-2}$ \citep{Boselli_et_al_2018}. This clearly demonstrates that the choice of methodology can lead to very different results, and that the surface brightness limit is not relevant if not computed in the exact same way for all surveys. In addition, this limit depends on how clean the images are, so we could have obtained slightly different values had we chosen other images.

Nonetheless, the re-computed surface brightness limit enables us to compare our surveys. As expected, CFIS is shallower than MATLAS. VESTIGE reaches a depth similar to MATLAS, contrary to what was indicated with the SB limits in the literature, but this is not surprising after inspection of the images. Indeed, we noted that VESTIGE images looked very similar to NGVS (that has a depth similar to MATLAS) and even better due to the absence of the prominent ghost reflections that polluted the older filter set of MegaCam, as discussed in Section \ref{fig:appendix-deep_images-ngvs_vestige}. We emphasize again that the important point for our study is not an ill-defined surface brightness limit value, but the fact that all images come from the same instrument and same reduction pipeline, which made them appear relatively similar. Furthermore, the SB limit is not at all synonym of detection limit of the LSB features \citep[e.g.,][]{Martin_et_al_2020,Roman_et_al_2020}.

\section{Galaxy sample}\label{section:appendix-galaxy_sample}
\begin{figure*}
  \centering
   \includegraphics[width=\linewidth]{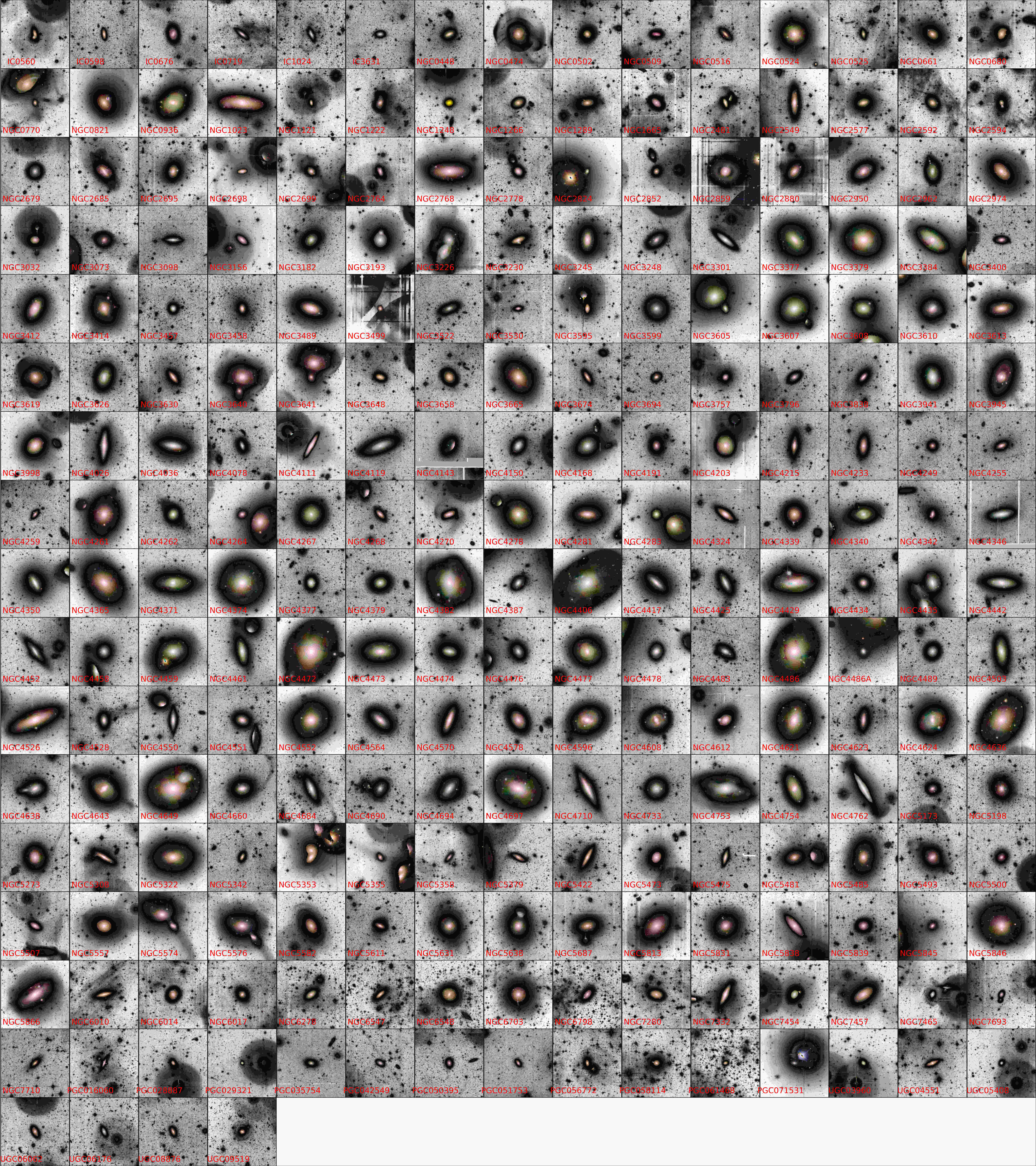}    
   \caption{Color-composite image of the ETGs in our final sample. The forefront true-color image is taken from PanSTARRS $r,g,i$ images and displays the galaxies as seen in shallow surveys. The greyscale image shows the extent of the galaxy as seen in deep CFHT observations. The field of view is 0.1 degree.}
    \label{fig:appendix-thumbnails_composite_ETGs}
\end{figure*}

\begin{figure*}
  \centering
   \includegraphics[width=\linewidth]{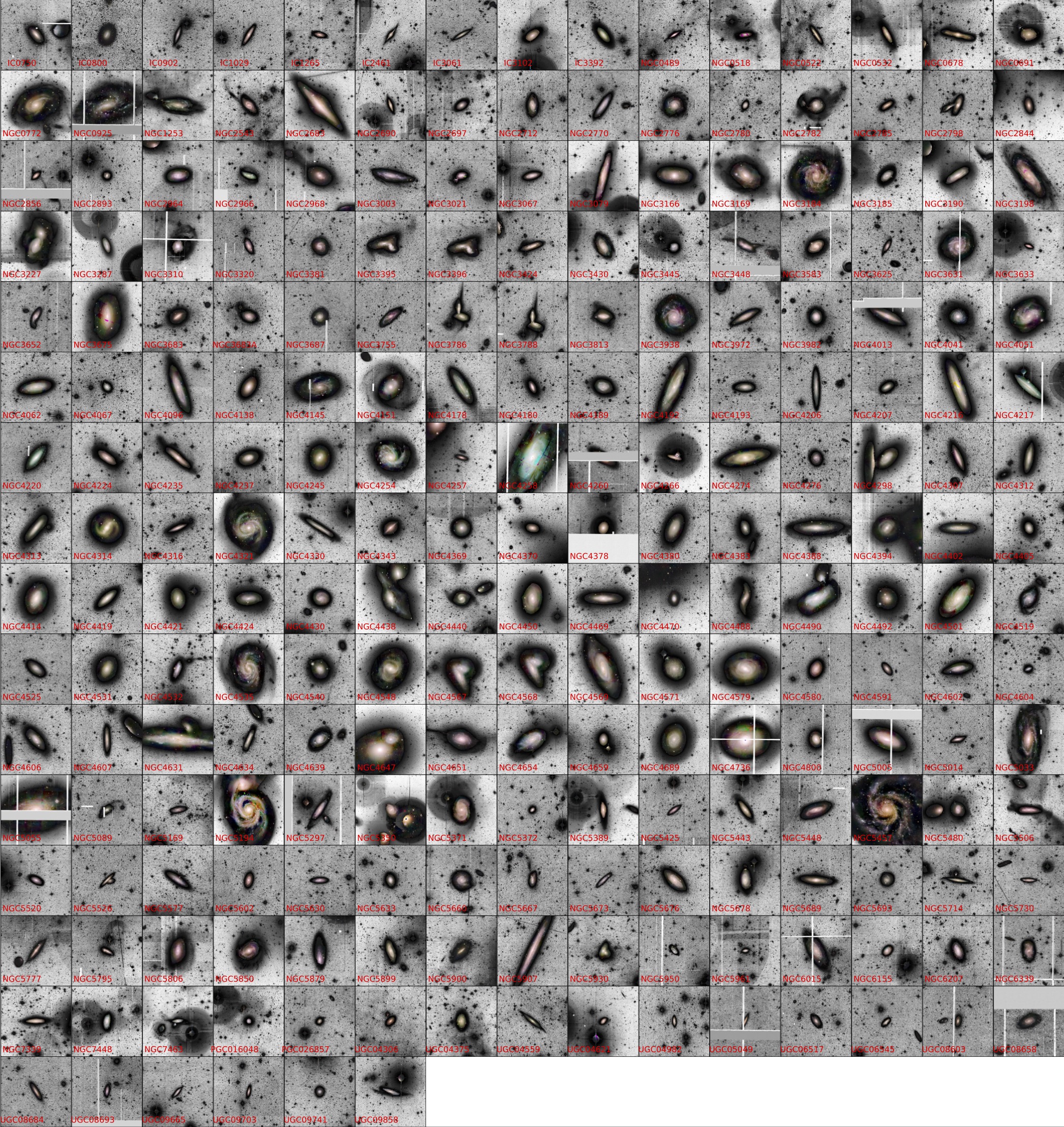}    
   \caption{Color-composite image of the LTGs in our final sample, alphabetically ordered. The forefront RGB image is taken from PanSTARRS $r,g,i$ images and displays the galaxies as seen in shallow surveys. The greyscale image shows the extent of the galaxy as seen in deep CFHT observations. The field of view is 0.1 degree.}
    \label{fig:appendix-thumbnails_composite_LTGs}
\end{figure*}

This section describes our sample of 475 galaxies. Table \ref{table:appendix-table_properties_galaxies} provides a description of some properties of the galaxies. Figures \ref{fig:appendix-thumbnails_composite_ETGs} and \ref{fig:appendix-thumbnails_composite_LTGs} display color-composite thumbnails generated from true-colour PanSTARRS images and greyscale deep CFHT images, for ETGs and LTGs, respectively.

The stellar mass is computed differently for ETGs and LTGs. 
Following \cite{Cappellari_et_al_2013a}, ETGs are characterised by their dynamical mass $M_{\rm JAM}$ estimated by Jeans anisotropic modelling (JAM) \citep{Cappellari_et_al_2008}. The relation between the stellar mass of ETGs $M_{\rm \star,ETG}$, the total galaxy luminosity $L$, the mass-to-light ratio $(M/L)_{\rm JAM}$ and dynamical mass $M_{\rm JAM}$ is given by $ M_{\rm JAM} \equiv L \times (M/L)_{\rm JAM}$ and $M_{\rm \star,ETG} \approx 0.87 M_{\rm JAM} $. The 0.87 factor linking stellar and dynamical masses comes from the estimated median fraction of dark matter mass of 13\% enclosed in $M_{\rm JAM}$ \citep{Cappellari_et_al_2013a}. $M_{\rm JAM}$ has the advantage of being more precise than stellar mass estimated from luminosity, as it does not require assumptions on the initial mass function, even though an assumption on the fraction of dark matter is needed. 
For LTGs, we did not have such $M_{\rm JAM}$ values, so we estimated their stellar mass $M_{\rm\star,LTG}$ from their $K$-band luminosity $L_K$ and the stellar mass-to-light ratio $(M/L)_K$, as $M_{\rm \star,LTG} = (M/L)_K \times L_K$, assuming a fixed $(M/L)_K = 0.8 M_\odot/L_\odot$ from \cite{Cappellari_et_al_2013}. The assumed $(M/L)_K$ ensures agreement between $M_{\rm JAM}$ and the $K$-band luminosity at the lowest masses \citep{Cappellari_et_al_2013}.

\begin{figure}
    \centering
    \includegraphics[width=0.35\linewidth]{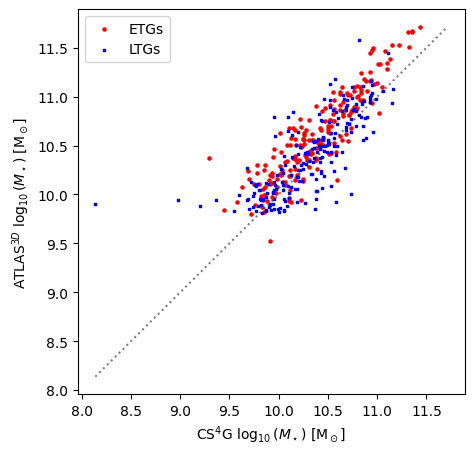}
    \caption{Comparison between our estimated stellar masses and the stellar masses from the CS$^4$G catalogue from \protect\cite{Sanchez-Alarcon_et_al_2025} for the 409 galaxies in common between our galaxy sample and the CS$^4$G catalogue. ETGs are plotted in red circles, LTGs in blue squares. The black dotted line represents the 1:1 line.} 
    \label{fig:appendix_mass_ours_S4G}
\end{figure}

As a sanity check, we compare our estimated stellar masses to the ones from the Complete Spitzer Survey of Stellar Structure in Galaxies (CS$^4$G) catalogue \citep{Sanchez-Alarcon_et_al_2025}. This catalogue contains the original S$^4$G galaxies \citep{Sheth_et_al_2010} with their stellar masses measured from \cite{Munoz-Matos_et_al_2015}, an extension to ETGs \citep{Watkins_et_al_2022} and additional disc galaxies. The 3239 CS$^4$G galaxies have consistent homogenised photometric parameters. We compare the stellar masses for the 409 galaxies we have in common in Figure \ref{fig:appendix_mass_ours_S4G}.
Overall, the stellar masses follow the 1:1 line, although some scatter is visible, comforting our approach. We seem to slightly overestimate the masses of the most massive galaxies (above 10$^{10.5}$-10$^{11}$ M$_\odot$ compared to CS$^4$G). As we are interested in the trends versus galaxy mass, rather than in the actual, precise mass values, we keep our mass estimates throughout the paper.

\begin{table*}
    \centering
    \caption{Sample of  galaxies studied in this work. The full table is available electronically  from the CDS databases.  Column (1): Galaxy name. Column (2): Right ascension (degree). Column (3): Declination (degree). Column (4): Morphological type. Column (5): Distance (in Mpc). Column (6): Stellar mass (in M$_\odot$), computed as described in Section \ref{section:results-effect_galaxy_mass}. Column (7): Projected half-light effective radius in arcsecond. Column (8): Environmental density $\rho_{10}$ (in Mpc$^{-3}$), i.e. mean density of galaxies inside a sphere centred on the galaxy and containing the 10 nearest neighbors. Column (9): Virgo cluster membership (1 if the galaxy is in Virgo, 0 otherwise). Column (10): Survey from which the image was taken. Column (11): Photometric bands available. Column (12): Small-scale environment of the galaxy, defined in Section \ref{section:results-effect_environment}: 1 if the galaxy is isolated, 0 if it is in pair (i.e. there is an annotated nearby companion susceptible to produce tidal interactions).  Values in columns (1),(2),(3),(4),(5),(7) and (9) are drawn from \protect\cite{Cappellari_et_al_2011}, while Column (8) is from \protect\cite{Cappellari_et_al_2011b}. When a galaxy was imaged by several surveys in different bands, they are all listed. However, all the annotations were made on the reference $r$-band image, and on the deepest available image (i.e., MATLAS rather than CFIS when both surveys were available). }
    \label{table:appendix-table_properties_galaxies}
    \begin{tabular}{c c c c c c c c c c c c}
        \hline 
        Galaxy & RA & Dec & Type & D& Mass & $\log_{10}(R_e)$ & $\log_{10}(\rho_{10})$ & Virgo & Survey & Bands & Isolated \\
        & [deg] & [deg] & & [Mpc] & [M$_\odot$] &  [$\arcsec$] & [Mpc$^{-3}$] & & & &\\
        (1) & (2)  &(3)  & (4)  & (5)  & (6)& (7) & (8)& (9) & (10) & (11) & (12)   \\
        \hline
        IC0560 & 146.472 & -0.269 & ETG & 27.2 & 9.81e+09 & 1.11 & -1.91 & 0 & MATLAS & rg & 1 \\
        IC0598 & 153.202 & 43.146 & ETG & 35.3 & 1.61e+10 & 1.02 & -2.31 & 0 & MATLAS & rg & 1 \\
        IC0676 & 168.166 & 9.056 & ETG & 24.6 & 1.41e+10 & 1.35 & -1.41 & 0 & MATLAS & rg & 1 \\
        IC0719 & 175.077 & 9.01 & ETG & 29.4 & 3.75e+10 & 1.1 & -1.91 & 0 & MATLAS & r & 0 \\
        IC0750 & 179.718 & 42.722 & LTG & 36.8 & 1.24e+11 & 1.24 & -2.29 & 0 & CFIS & r & 0 \\
        \vdots & \vdots & \vdots & \vdots & \vdots & \vdots & \vdots & \vdots & \vdots & \vdots & \vdots & \vdots  \\
        UGC09519 & 221.588 & 34.371 & ETG & 27.6 & 1.01e+10 & 0.87 & -2.45 & 0 & MATLAS & rg & 1 \\
        UGC09665 & 225.385 & 48.32 & LTG & 40.3 & 2.08e+10 & 1.11 & -1.71 & 0 & CFIS & r & 0 \\
        UGC09703 & 226.359 & 46.565 & LTG & 39.1 & 9.27e+09 & 1.07 & -1.59 & 0 & CFIS & r & 1 \\
        UGC09741 & 227.14 & 52.296 & LTG & 39.6 & 7.43e+09 & 1.16 & -1.98 & 0 & CFIS & r & 1 \\
        UGC09858 & 231.673 & 40.564 & LTG & 40.9 & 2.93e+10 & 1.24 & -2.33 & 0 & CFIS & r & 1 \\
        \hline
    \end{tabular}
\end{table*}

\clearpage

\section{Individual features}\label{section:appendix-coordinates_features}
In Table \ref{table:appendix-nb_features}, we provide the number of tidal features identified per galaxy, while in Table \ref{table:appendix-list_features}, we provide the list of the individual 199 tidal tails and 100 streams that were kept after our selection process. 
\begin{table}
\centering
\caption{Number of tidal features per galaxy, as computed in Section \ref{section:results-census_features}. The full table is available electronically at CDS databases. Column (1): Galaxy name. Column (2): Number of tidal tails. Column (3): Number of streams. Column (4): Number of shells.}
\label{table:appendix-nb_features}
\begin{tabular}{ccccccccc}
\hline
Galaxy & \# tails & \# streams & \# shells \\
(1) & (2)  &(3)  & (4)  \\
\hline
IC0560 & 0 & 0 & 0.0 \\
IC0598 & 0 & 0 & 0.0 \\
IC0676 & 0 & 0 & 0.0 \\
IC0719 & 0 & 0 & 0.0 \\
\vdots & \vdots & \vdots & \vdots \\
UGC09665 & 0 & 0 & 0.0 \\
UGC09703 & 0 & 0 & 0.0 \\
UGC09741 & 0 & 0 & 0.0 \\
UGC09858 & 2 & 0 & 0.0 \\
\hline
\end{tabular}
\end{table}

\LTcapwidth=\linewidth
\begin{table}
\centering
\caption{The list of the individual annotated tidal tails and streams. The full table is available electronically at CDS databases. Column (1): identification number of the feature. Column (2): Feature type.  Column (3): Galaxy name. Column (4): Length (kpc). Column (5): Area (kpc$^2$). Column (6): Width (defined as the area divided by the length). Column (7): Median surface brightness (mag$\,$arcsec$^{-2}$) inside the annotation. Column (8): Percentage of the total galaxy flux contained in that feature. Column (9): Total galaxy magnitude (mag, defined as $-2.5\times \log_{10}($total\_galaxy\_flux$) + 30$)), used to normalise Column (8). }
\label{table:appendix-list_features}
\begin{tabular}{c c c c c c c c c}
\hline
\footnotesize
Feature \# & Type  & Galaxy & Length & Area & Width & SB & Flux & Total galaxy magnitude \\ 
& & & [kpc] &[kpc$^2$] & [kpc] & [mag$\,$arcsec$^{-2}$] & [\%] & [mag]\\
(1) & (2)  &(3)  & (4)  & (5)  & (6)& (7) & (8) & (9) \\
\hline
1 & Stream & IC1024 & 49.4 & 270.8 & 5.5 & 27.3 & 2.6 & 12.6 \\
2 & Stream & IC1024 & 32.4 & 126.5 & 3.9 & 27.5 & 1.0 & 12.6 \\
3 & Stream & IC3102 & 25.9 & 113.0 & 4.4 & 26.2 & 2.7 & 11.1 \\
4 & Stream & NGC0474 & 135.1 & 751.6 & 5.6 & 25.3 & 21.6 & 10.9 \\
\vdots & \vdots & \vdots & \vdots &\vdots & \vdots & \vdots & \vdots & \vdots \\
296 & Tail & UGC08693 & 11.3 & 39.7 & 3.5 & 26.5 & 0.7 & 13.6 \\
297 & Tail & UGC08693 & 15.9 & 34.4 & 2.2 & 25.4 & 2.2 & 13.6 \\
298 & Tail & UGC09858 & 32.2 & 160.8 & 5.0 & 25.1 & 2.5 & 11.7 \\
299 & Tail & UGC09858 & 26.4 & 104.0 & 3.9 & 25.3 & 3.7 & 11.7 \\
\hline  
\end{tabular}
\end{table}

To ensure full reproducibility of our results, and as a complement to Table \ref{table:appendix-list_features}, we provide a file in the DS9 Region format that contains the coordinates of our LSB features: tails, streams but also shells, haloes and inner galaxies. An truncated example is shown below, while the full version is available electronically in the CDS databases. The file lists the identification number of the feature, the host galaxy, the type of feature and the coordinates of its contours, which are given in the format \textit{polygon(ra$_1$,dec$_1$,ra$_2$,dec$_2$,...,ra$_n$,dec$_n$)}. This region file can be opened with SAOImageDS9 and Aladin. Arbitrary colours were associated to each feature type, e.g., red for \textit{Stream}, blue for \textit{Tidal Tails}, orange for \textit{Shells}, green for \textit{Haloes} and magenta for \textit{Inner Galaxy}.

\begin{verbatim}
# Region file format: DS9 version 4.1
global color=yellow dashlist=8 3 width=1 font="helvetica 10 normal roman" select=1 
highlite=1 dash=0 fixed=0 edit=1 move=1 delete=1 include=1 source=1
icrs
# Feature0 IC1024 Streams
polygon(217.859,3.02384,...,217.872,3.02806) # color=red
# Feature1 IC1024 Streams
polygon(217.849,3.05411,...,217.853,3.05204) # color=red
# Feature2 IC3102 Streams
polygon(184.33315,6.7513204,...,184.32008,6.7470474) # color=red
# Feature3 NGC0474 Streams
polygon(20.0919,3.56068,...,20.0848,3.55207) # color=red
# Feature4 NGC0474 Streams
polygon(20.034,3.34547,...,20.0243,3.34832) # color=red

\end{verbatim}

\clearpage
\section{Masks, background subtraction and luminosity} \label{section:appendix_flux_estimation}
\begin{figure*}
  \centering
  \includegraphics[width=0.8\linewidth]{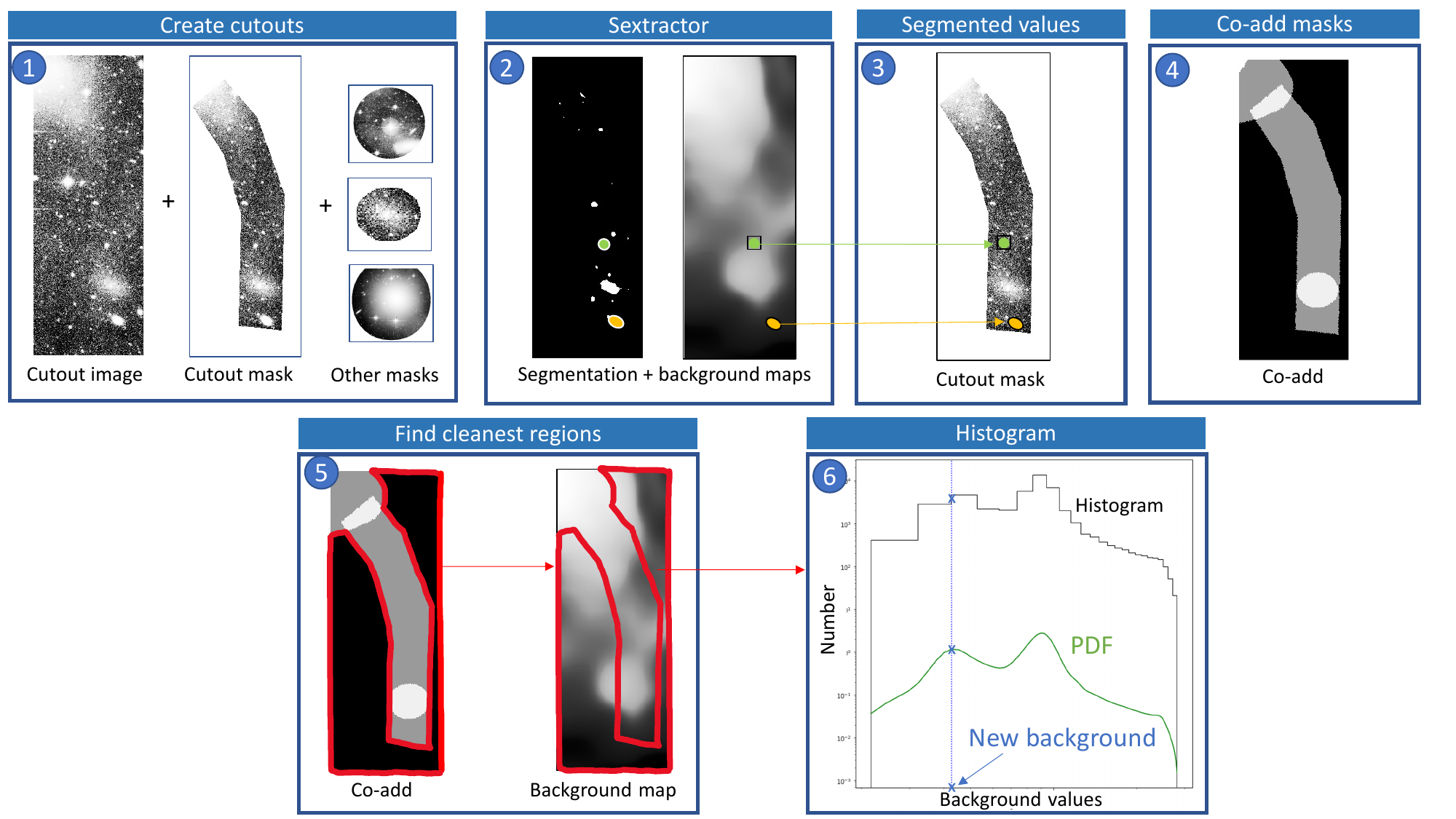}
      \caption{Illustration of the refined luminosity estimation process, for one stream feature around one galaxy. \textit{Step 1}: the cutouts of the mask and image of the feature of interest are created, as well as masks of other annotations around the galaxy.\textit{ Step 2}: source extraction and background measurement are performed using SExtractor. Two segmented regions are highlighted in orange and green, to illustrate how the pixels will be replaced in the next step. \textit{Step 3}: segmented pixels are replaced by the corresponding background map values. Two examples of such segmented regions are represented in green and orange. \textit{Step 4}: the masks of the other annotations are co-added on the same field of view as the feature of interest. \textit{Step 5}: the contours of the cleanest regions, with the least number of annotations overlapping, are retrieved from the co-add and applied to the background map. Here, the cleanest regions are represented inside the red contour. \textit{Step 6}: inside the cleanest regions, the histogram of the background map values is created (in black). Then, the PDF is estimated (in green), and the peaks of the PDF are computed. Only the first significant peak is kept (in blue), and its value is taken as the new background value. }
         \label{fig:mask_refined}
\end{figure*}

In this section, we describe the method we used to obtain a better estimate of the luminosity and the SB of a structure.The process for one feature annotated by one contributor for a specific galaxy is divided in several steps, each of which are described below. An illustration of this process is presented in Figure \ref{fig:mask_refined}.

The process starts with the creation of two cutouts (step 1): a mask showing only the feature (i.e. blank values outside the feature), and an image of the feature with its surroundings (i.e. a cropped version of the original image). Working with cutouts reduces the size of the files and the computation time required to process them. It must be noted that in the particular case of haloes, we only retained the region between the external border of the main galaxy and the external border of the halo, to remove the influence of the centre of the galaxy. Hence, our halo masks have `donut' shapes when they are the feature of interest. 

Then, mask cutouts of other annotations for the same contributor and the same galaxy are created: the entire halo\footnote{We use the term `entire halo' to refer to the annotations that retain both the inner part of the main galaxy and the donut-shaped outer halo.}, companion galaxies (including small dwarfs) and ghosted reflections. These annotations are important, as they might overlap with the feature of interest.

The next step is source extraction to remove bright foreground sources (step 2). The software SExtractor \citep{Bertin_and_Arnouts_1996} was run in double mode on the feature of interest. The segmentation was done on the mask cutout, while the background measurement was done on the cutout image\footnote{After some tests, we used the following threshold for detection of tidal features: DETECT\_THRESH$=$2 and DETECT\_THRESH$=$12 for haloes. To estimate the background, we use the parameters BACK\_PHOTOTYPE$=$Local and BACK\_SIZE$=$32.}.

The pixels masked on the segmentation map are then replaced by the corresponding values from the background map, and then applied to the mask cutout (step 3). The last part of this step is a cleaning one. Indeed, there may be erroneous pixels in the mask cutout, due to artefacts from the instrument (like CCD gaps). After inspection on some initial test cases, these erroneous pixels have large negative values, so all the pixels with a value smaller than a given threshold are replaced by the corresponding pixels in the background map.

We re-estimate the background after the pixel replacement (steps 4 - 6). This step is important, as it enables a more precise background estimation, which is crucial when dealing with LSB features. The images are supposed to have a flat background equal to zero after being processed by Elixir-LSB. However, there may be local variations, due to contamination sources like bright stars or cirrus, that locally cause non-zero background values. We want to remove the influence of other structures, i.e. other annotations that may be overlapping with the feature of interest.

The re-estimation of the background starts with the co-addition of all the cutouts masks of the other annotations for that contributor and that galaxy, that were previously created during step 1. They are co-added and re-projected\footnote{The re-projection was done using the \textit{reproject} Python package, \href{https://reproject.readthedocs.io/en/stable/}{https://reproject.readthedocs.io/en/stable/}} to the same field of view as the feature of interest (step 4). This co-add is applied on SExtractor's background map. The cleanest regions, i.e. where there are no footprint of any annotation, are retrieved (step 5). In the case where there are annotations everywhere, the regions where there are the smallest number of overlapping annotations are kept. The new background is then evaluated in these cleanest regions. To that end, the histogram of the background map values in the cleanest regions is created (step 6), and the corresponding probability distribution function (PDF) is obtained from a kernel density estimation. As the PDF can be multimodal, it is not correct to simply keep the mean or mean value, nor the highest or smallest peak of the PDF. After some tests, we decided to use the first significant peak of the PDF as the new background value. A peak is considered significant if the computation of its prominence is higher than a given threshold. 

Finally, the newly estimated background value is subtracted from all pixels in the cutout mask; at the end of this step the final cutout mask is obtained. From this final cutout mask, the flux $f$ can be retrieved simply by summing all the pixels inside it. The luminosity $L$ is obtained through the formula:
\begin{equation}
    L = 4\pi\times D^2 \times f
\end{equation}
where $D$ is the distance of the galaxy and $f$ the flux.
Likewise, the SB can be computed from that final mask.

\begin{figure*}
  \centering
  \includegraphics[width=0.75\linewidth]{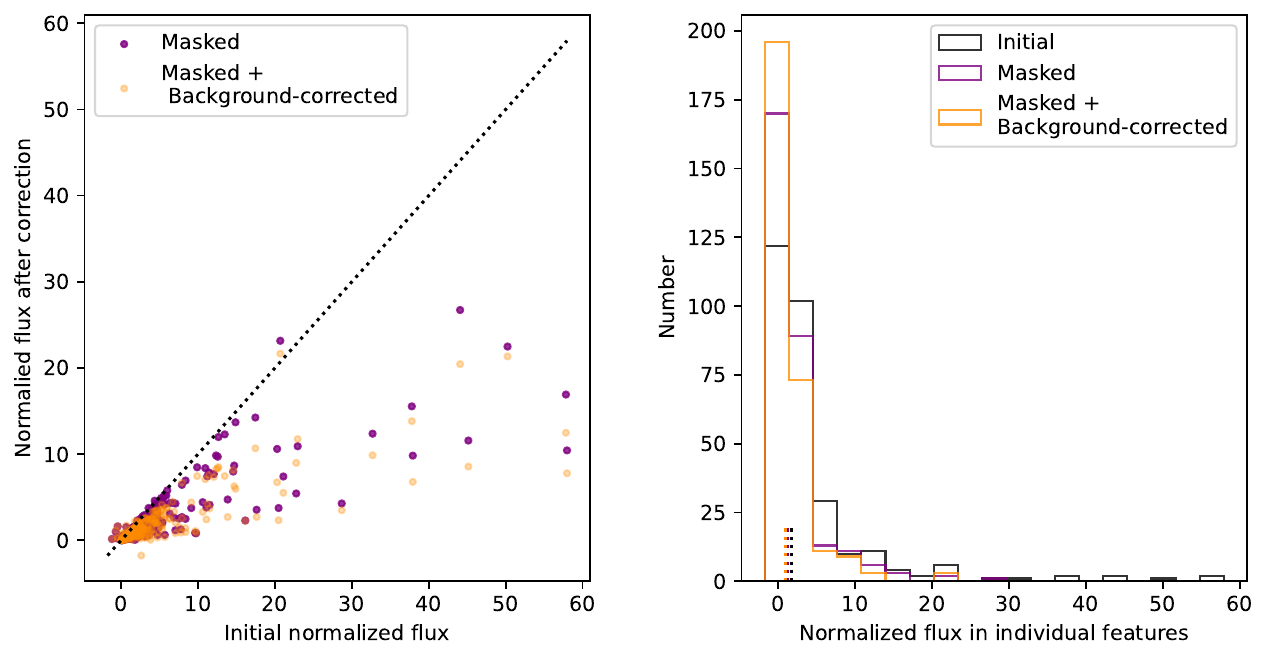}
      \caption{Comparison of the fraction of flux in individual tidal features (tails and streams), normalized with respect to the flux of the entire annotated halo. This fraction is computed on the initial cutout images, on the masked (i.e. bright sources masked) and on the final background-corrected (i.e. bright sources masked and background correction) images. \textit{Left}: Scatter plot of the fraction of initial versus masked (in purple) and background-corrected (in orange) fluxes for each individual feature. \textit{Right}: Histogram of the fraction of flux in tidal features, computed on the initial (in black), masked (in purple) and final background-corrected (in black) images. }
    \label{fig:appendix-effect_bkcorrection_flux}
\end{figure*}

We now investigate the impact that masking of the bright sources and of our background correction can have on the retrieved flux of the features. To that end, we compare the flux in features in the initial (i.e., not masked, not background-corrected), masked (not background-corrected) and final (masked, background-corrected) cutout images. We normalise the flux in features (individual tails and streams) to the flux of the whole halo (i.e. without tidal features). Figure \ref{fig:appendix-effect_bkcorrection_flux} presents the comparison.

It appears that the most crucial step of our method is masking the bright stars rather than the re-estimation of the background. Bright sources account for 67\% of the flux of tidal features, while an additional 33\% comes from not performing the background re-estimation.  Indeed, the median (resp. mean) flux fraction in individual tidal features is 1.8\% (resp. 4.5\%) for the initial images; against 1.2\% (resp. 2.4\%) for the clean images and 0.9\% (resp. 1.8\%) for the background-corrected images.

\clearpage
\section{Trends as a function of the morphological type} \label{appendix:trends_vs_morphtype}

\begin{figure*}
  \centering
  \includegraphics[width=0.4\linewidth]{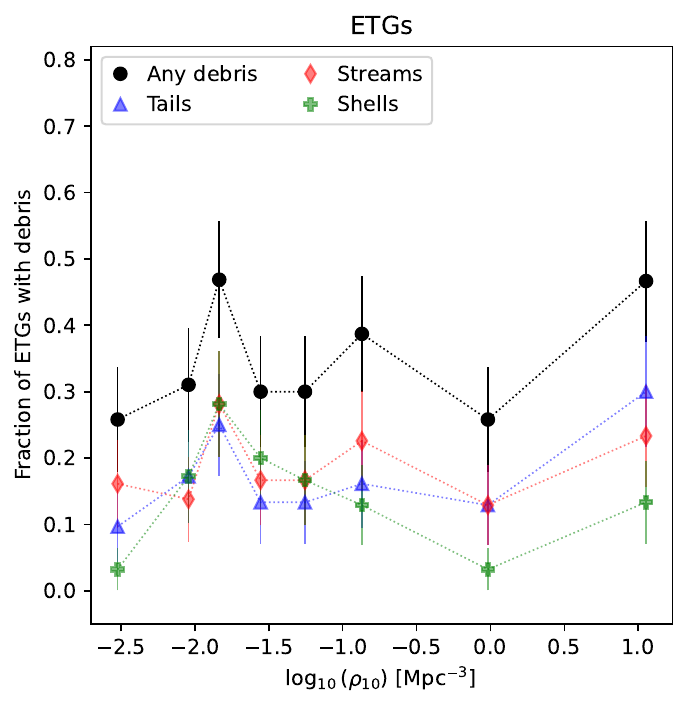}
  \hspace{10cm}
  \vspace{0.1cm}
   \includegraphics[width=0.3\linewidth]{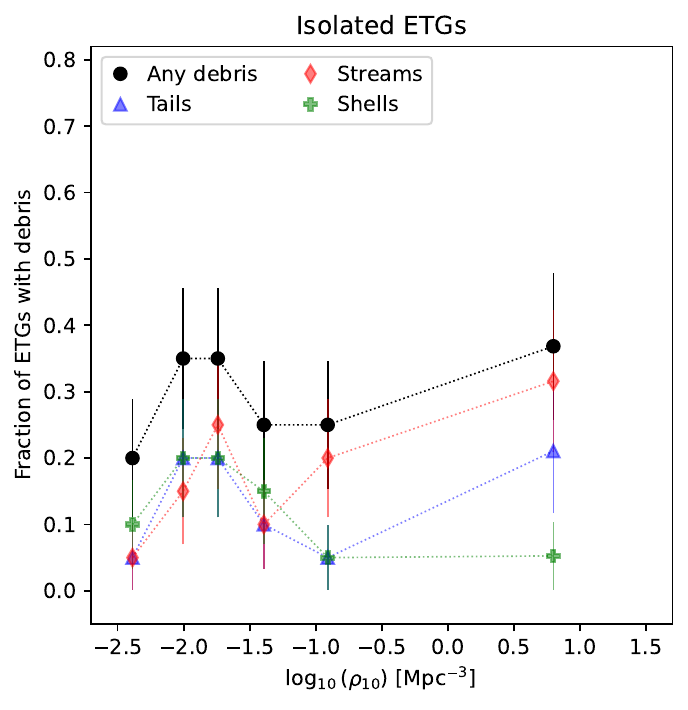}    
    \hspace{0.1cm}
   \includegraphics[width=0.3\linewidth]{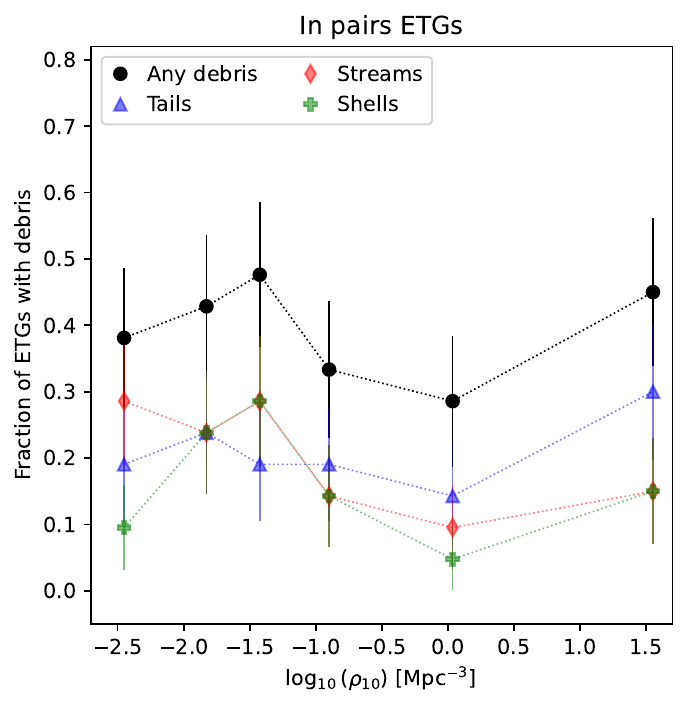}    
   \caption{Fraction of ETGs that have tidal features as a function of the galaxy density $\rho_{10}$ (in Mpc$^{-3}$), per bin of $\rho_{10}$. Each bin contains approximately the same number of galaxies (i.e. about 30 galaxies per bin for the top plot, 20 for the bottom left and 21 for the bottom right). The fraction of ETGs hosting any type of debris is plotted in black, ETGs hosting tails in blue, streams in red and shells in green. The error bars represent the standard deviation on proportions in each bin. Galaxies in the Virgo cluster have $\log_{10}(\rho_{10}) > -0.4$. \textit{Top}: Fraction of ETGs with debris considering all ETGs together. \textit{Bottom left}: Fraction of ETGs with debris only considering isolated (i.e. post-merger) ETGs (see Section \ref{section:results-effect_environment}). \textit{Bottom right}: Fraction of ETGs with debris only considering  ETGs in pairs, i.e. that could be undergoing tidal interactions with a massive companion.}
    \label{fig:appendix-ETGs_fraction_galdebris_vs_env}
\end{figure*}

\begin{figure*}
  \centering
  \includegraphics[width=0.4\linewidth]{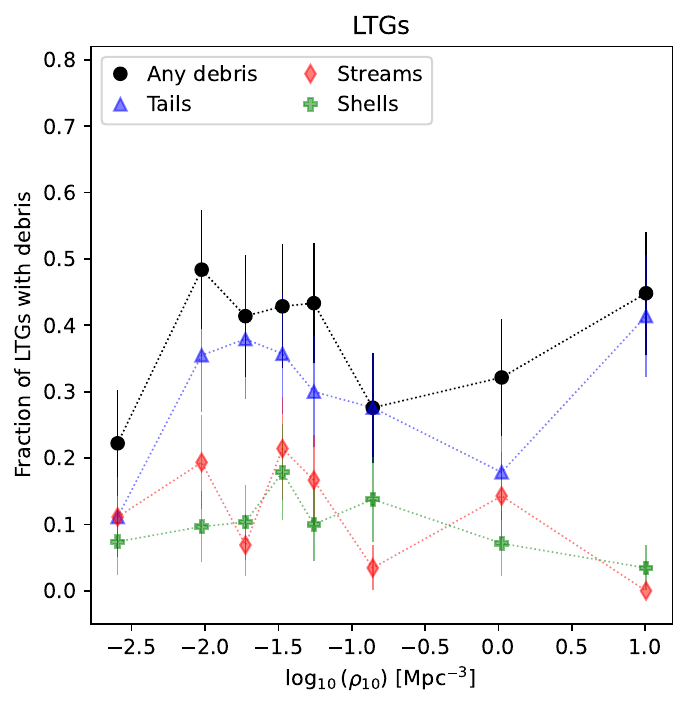}
  \hspace{10cm}
  \vspace{0.1cm}
   \includegraphics[width=0.3\linewidth]{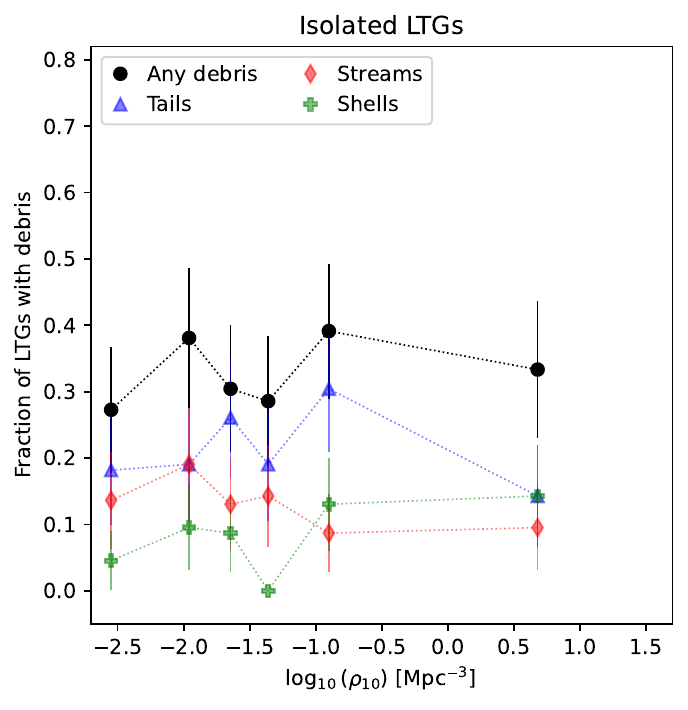}    
    \hspace{0.1cm}
   \includegraphics[width=0.3\linewidth]{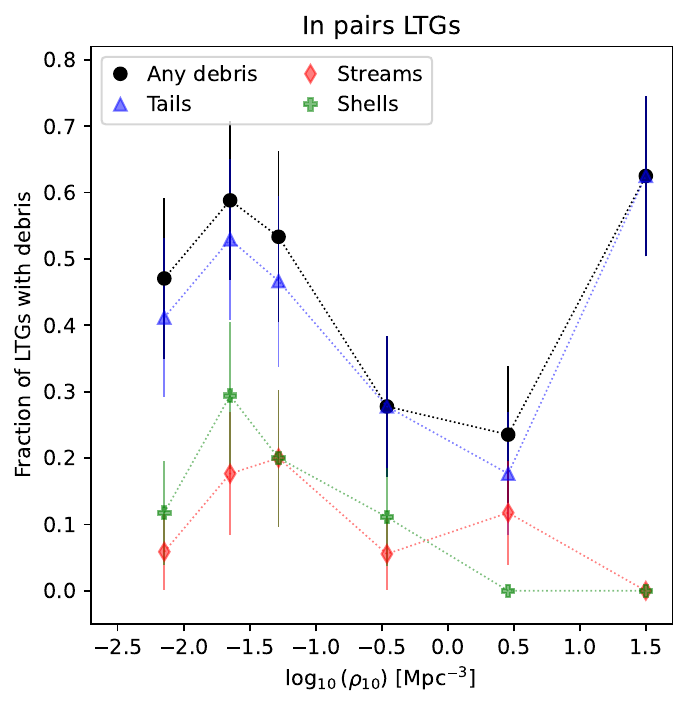}    
   \caption{Fraction of LTGs that have tidal features as a function of the galaxy density $\rho_{10}$ (in Mpc$^{-3}$), per bin of $\rho_{10}$. Each bin contains approximately the same number of galaxies (i.e. about 29 galaxies per bin for the top plot, 21 for the bottom left and 17 for the bottom right). The fraction of LTGs hosting any type of debris is plotted in black, LTGs hosting tails in blue, streams in red and shells in green. The error bars represent the standard deviation on proportions in each bin. Galaxies in the Virgo cluster have $\log_{10}(\rho_{10}) > -0.4$. \textit{Top}: Fraction of LTGs with debris considering all LTGs together. \textit{Bottom left}: Fraction of LTGs with debris only considering isolated (i.e. post-merger) LTGs (see Section \ref{section:results-effect_environment}). \textit{Bottom right}: Fraction of LTGs with debris only considering LTGs in pairs, i.e. that could be undergoing tidal interactions with a massive companion.}
    \label{fig:appendix-LTGs_fraction_galdebris_vs_env}
\end{figure*}

\begin{figure*}
  \centering
   \includegraphics[width=0.35\linewidth]{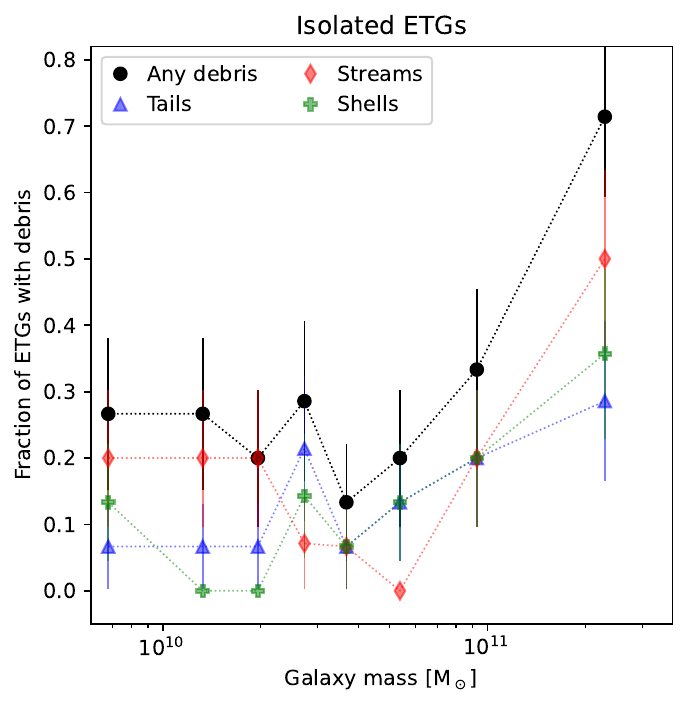}    
    \hspace{0.1cm}
   \includegraphics[width=0.35\linewidth]{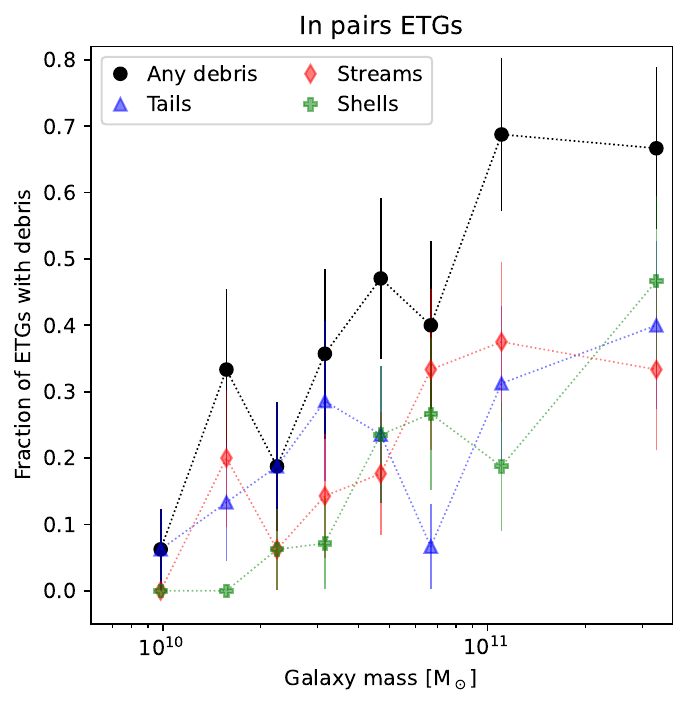}   
  \hspace{5cm}
  \vspace{0.1cm}
   \includegraphics[width=0.35\linewidth]{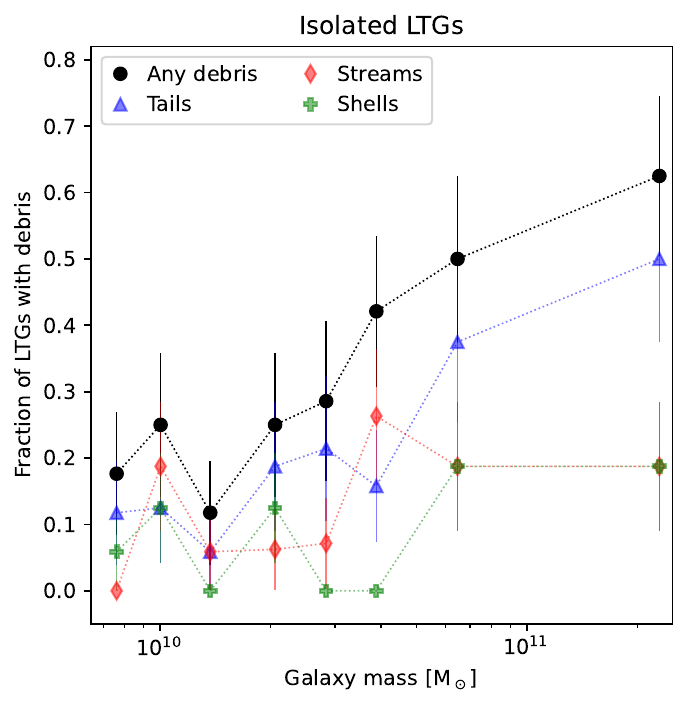}    
    \hspace{0.1cm}
   \includegraphics[width=0.35\linewidth]{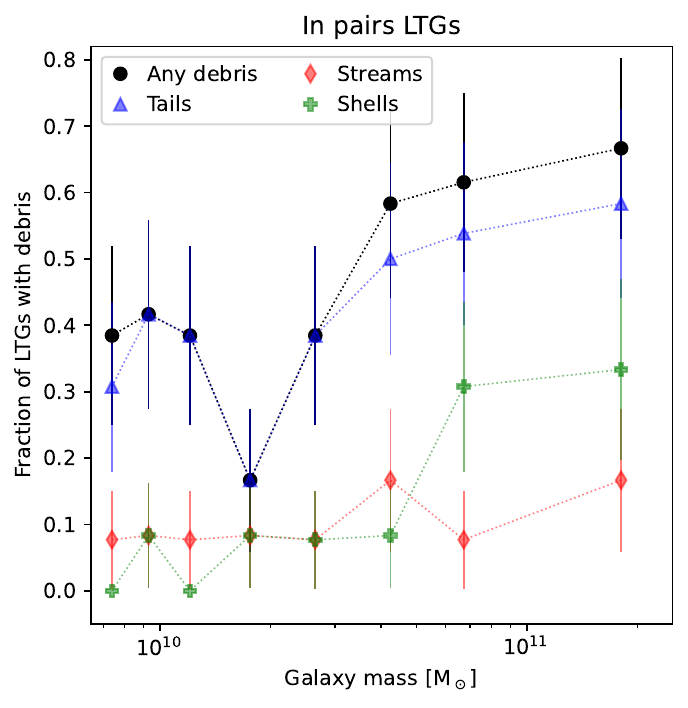}    
   \caption{Fraction of ETGs (\textit{top panel}) and LTGs (\textit{bottom panel}) that have tidal features as a function of the galaxy stellar mass (in M$_\odot$), per mass bin. Each bin contains approximately the same number of galaxies (i.e. about 15 galaxies per bin for the top plots, 16 for the bottom left and 12 for the bottom right plot). The fraction of galaxies hosting any type of debris is plotted in black, galaxies hosting tails, in blue, streams in red and shells in green. The error bars represent the standard deviation on proportions in each bin. \textit{Left column}: Fraction of galaxies with debris only considering isolated (i.e. post-merger) galaxies (see Section \ref{section:results-effect_environment}). \textit{Right column}: Fraction of galaxies with debris only considering  galaxies in pairs, i.e. that could be undergoing tidal interactions with a massive companion.}
    \label{fig:appendix-ETGs_LTGs_fraction_galdebris_vs_mass}
\end{figure*}

We complement the analyses of tidal features and stellar haloes by studying the trends as a function of the morphological type of the host galaxy, to determine whether this effect could be dominant. We investigate the fraction of galaxies hosting debris as a function of environmental density, separating post-merger galaxies from the ones in pairs, for ETGs and LTGs in Figures \ref{fig:appendix-ETGs_fraction_galdebris_vs_env} and \ref{fig:appendix-LTGs_fraction_galdebris_vs_env}. These plots are complementary to Figure \ref{fig:results-fraction_galdebris_vs_env}, with similar conclusions. We find hints of tidal feature generation in the group environment, with increased fractions of galaxies hosting all types of debris. Isolated galaxies do not show any prominent peak for the incidence of tidal features.
Similar plots are shown in Figure  \ref{fig:appendix-ETGs_LTGs_fraction_galdebris_vs_mass} for the fraction of galaxies hosting debris as a function of galaxy mass. As in Figure \ref{fig:results-fraction_galdebris_vs_mass}, we see an increase of the fraction of galaxies hosting debris as a function of galaxy mass. 

\clearpage

\clearpage
\section{Impact of the change of the set of filters} \label{section:appendix-filters_set}
\begin{figure}
    \centering
    \includegraphics[width=0.5\textwidth]{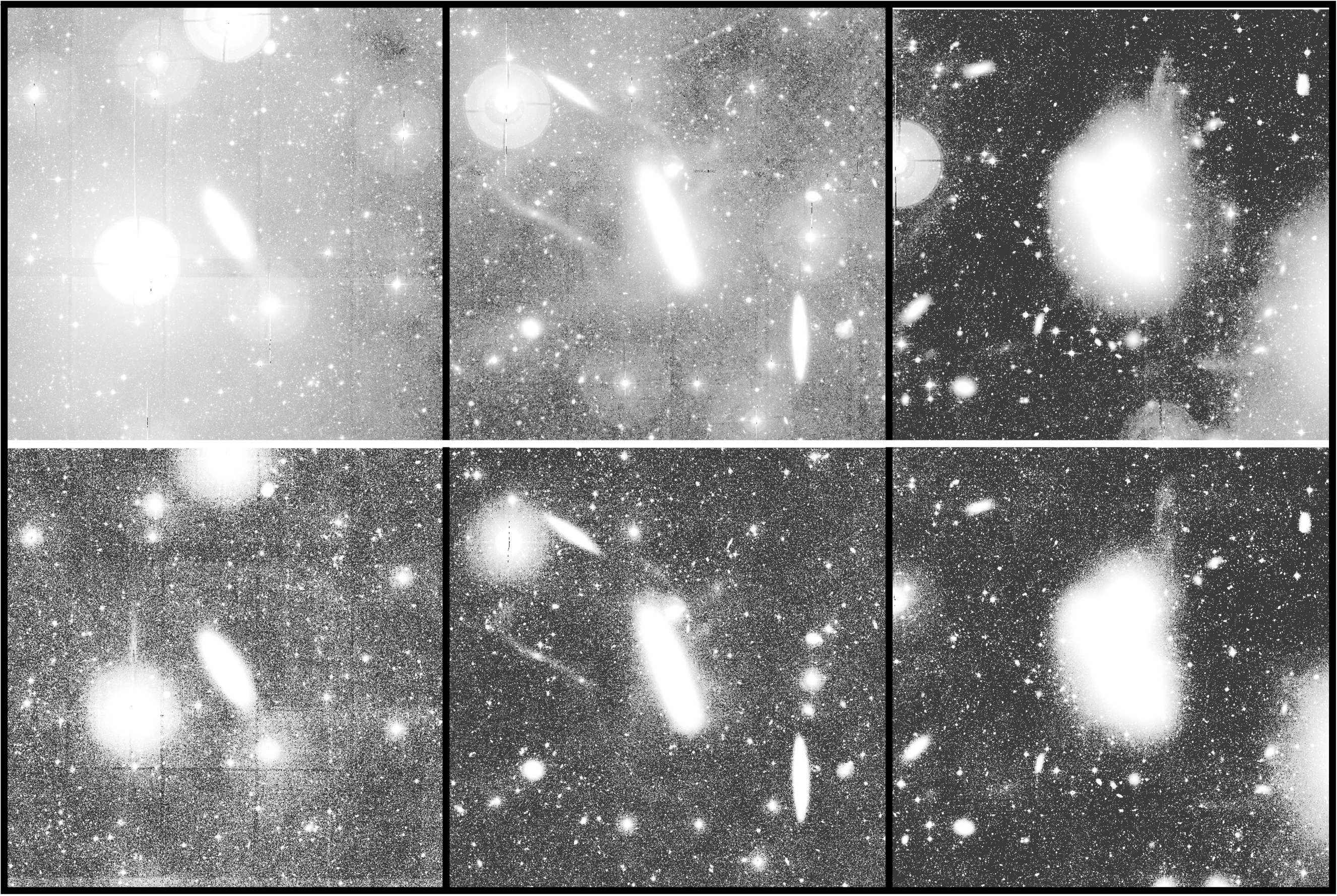}
    \caption{Comparison between images of the same galaxies from NGVS $g$-band images (\textit{top row}) and VESTIGE $r$-band images (\textit{bottom row}), with the same stretch and size. From left to right, NGC4178, NGC4216, and NGC4438.}
    \label{fig:appendix-deep_images-ngvs_vestige}
\end{figure}

The filter set of the MegaCam camera was changed after 2014. The old filters were in place for MATLAS and NGVS, whereas for CFIS and VESTIGE the new filters were used. This has an important impact on the images that we assessed in this work. Indeed, the properties of these new broad-band $ugriz$ filters differ from the old ones, with improved transmission properties (spatial and spectral), field of view  and different central wavelengths \citep{megacam_photometry}. 
This change has an impact on the shape and prominence of the ghost reflections as they depend on the filter. In Figure \ref{fig:appendix-deep_images-ngvs_vestige} we illustrate the fact that ghost reflections are much more prominent for the old MegaCam filters (MATLAS and NGVS) than for the new ones (VESTIGE and CFIS). 

\begin{figure*}
  \centering
  \includegraphics[width=0.75\linewidth]{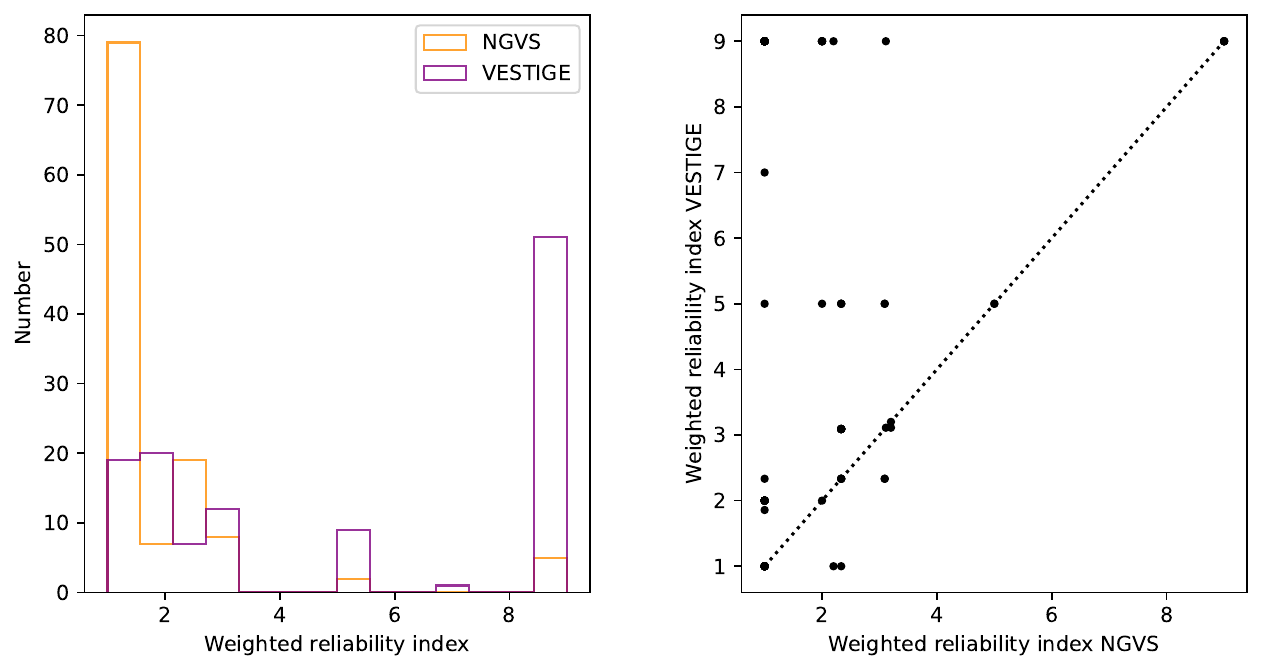} 
      \caption{\textit{Left:} Histogram of the weighted reliability index for NGVS (in orange) and VESTIGE (in purple). The higher the index, the cleaner the image. \textit{Right:} Scatter plot of the weighted reliability index of VESTIGE as a function of the one for NGVS for the same fields.}
    \label{fig:appendix-ngvs_vestige_reliability_index}
\end{figure*}

To assess this difference in a more quantitative way, we compared the cleanness of images of the same galaxies between NGVS ($g$-band, old filter) and VESTIGE ($r$-band, new filter)\footnote{We compare these two bands as VESTIGE only has the $r$-band available and NGVS has very few $r$-band images.}. We used the weighted reliability index defined in \cite{Sola_et_al_2022}; this number is based on the percentage of intersection between the annotation of the halo of a galaxy and annotations of sources of pollution (such as cirrus, bright ghost reflections, satellite trails). The higher the index, the cleaner the image. In Figure \ref{fig:appendix-ngvs_vestige_reliability_index}, we compare the weighted reliability index for the same galaxies annotated by a single contributor to have consistent annotations.
It clearly appears that NGVS images are much more impacted by contamination sources such as ghosted haloes and bright stars; 72\% of images have an index lower than 2. On the contrary, 43\% of VESTIGE images are clear of any contamination source, and 33\% have an index lower than 2. The new $r$-band filter images are much cleaner than NGVS, although $g$-band images are less affected by PSF contamination (i.e. less bright ghost reflections) than the $r$-band. 
The identification of tidal features is therefore on average easier in VESTIGE images, even though VESTIGE is slightly less deep than NGVS. This shows that instrumental effects including internal reflections due to filters (new versus old) have almost as much of an impact as the depth of the survey in the detection and identification of tidal features.

\begin{figure*}
    \centering
    \includegraphics[width=0.75\linewidth]{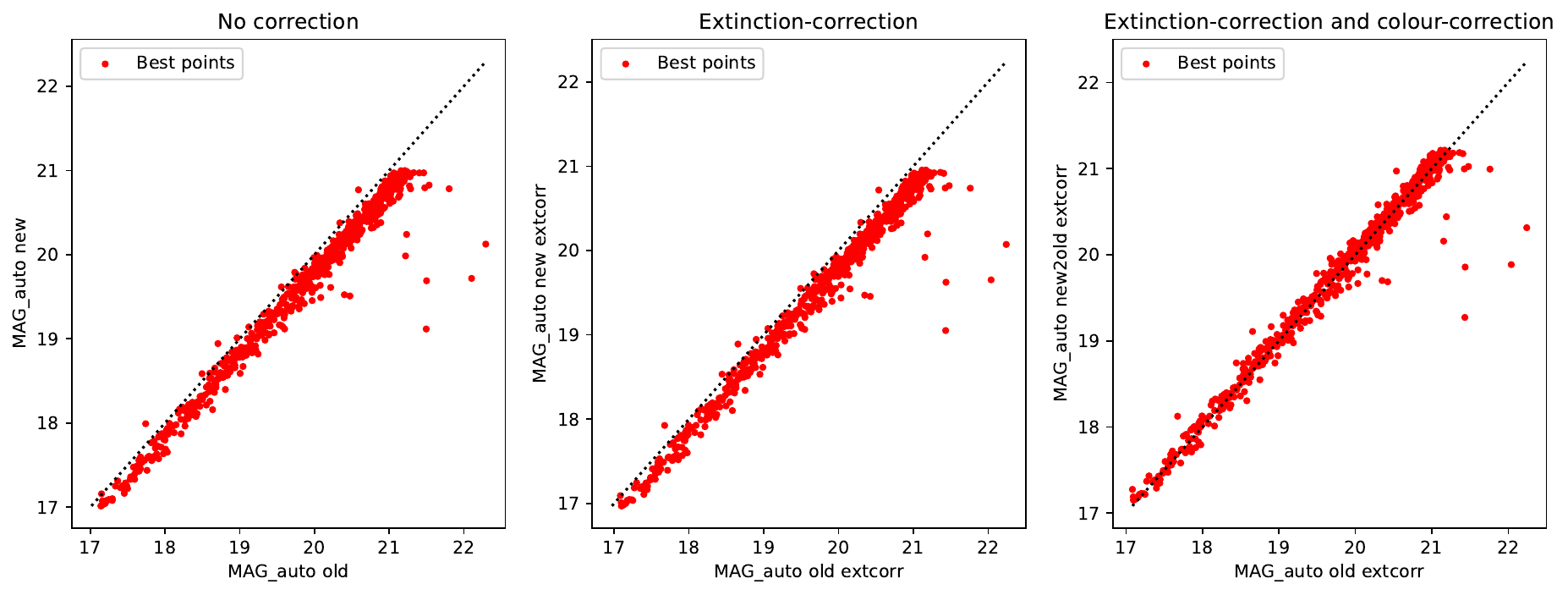}
    \caption{Estimation of the colour-correction between MegaCam old and new $r$-band filters, and effect of extinction correction. Magnitudes are expressed in mag. \textit{Left}: Scatter plot of the magnitude in the new filter (MAG\_auto new) as a function of the old filter (MAG\_auto old) without extinction correction. \textit{Middle}: Scatter plot of the magnitude in the new filter (MAG\_auto new extcorr) as a function of the old filter (MAG\_auto old extcorr) with extinction correction. \textit{Right}: Scatter plot of the magnitude in the new filter for which the colour-correction has been applied (MAG\_auto new2old extcorr) as a function of the old filter (MAG\_auto old extcorr) with extinction correction and colour-correction.}
    \label{fig:appendix-color_correction}
\end{figure*}

We assessed the impact of the change of filter on the colour estimation. We selected four common fields in the MATLAS/CFIS and NGVS/VESTIGE surveys that were observed with the old and new $r$-band filter\footnote{Only a few NGVS $r$-band images were available.}.
We ran SExtractor on each field to automatically compute the segmentation maps, and we applied the same segmentation between the old and the new filters. We then performed a cross-match of the sources in order to select only the same regions in each pair of fields. From this, we only kept the segmented regions that were the cleanest, i.e. with a high enough signal-to-noise ratio, and which followed a relatively tight relation for the magnitude between the old and new filters. We called these kept `best points' and we present their flux in the old and new filters in the left panel of Figure \ref{fig:appendix-color_correction}.
A clear offset between the new and the old magnitudes is observed. 
It is not due to the  variation of extinction in the 4 different fields, as shown in the middle panel of Figure \ref{fig:appendix-color_correction}.    

Fitting a linear regression in order to convert the magnitudes from the new filter to the old filter,  we derived the following conversion formula:
\begin{equation}
    mag_{\rm r, new2old} = 1.019 \times mag_{\rm r, old} - 0.147 
\end{equation}
where $mag_{\rm r,new2old}$ is the extinction and colour-corrected magnitude in the $r$-band new filter converted to be in the old filter, and $mag_{\rm r, old}$ is the magnitude in the old filter. The `extinction and colour-corrected magnitude' corresponds to the magnitude of the object when corrected both for extinction and for the change of MegaCam filters which required colour-correction. The effect of this correction is shown in the right panel of Figure \ref{fig:appendix-color_correction}.


\bsp	
\label{lastpage}
\end{document}